\documentclass{ws-mpla}

\begin{document}

\lefthyphenmin=5
\righthyphenmin=5

\newcommand{\dzero}     {D\O}
\newcommand{\ppbar}     {\mbox{$p\bar{p}$}}
\newcommand{\ttbar}     {\mbox{$t\bar{t}$}}
\newcommand{\bbbar}     {\mbox{$b\bar{b}$}}
\newcommand{\ccbar}     {\mbox{$c\bar{c}$}}
\newcommand{\rar}       {\rightarrow}
\newcommand{\met}       {\mbox{$\not\!\!E_T$}}

\markboth{A.P. Heinson}
{Observation of Single Top Quark Production}

\catchline{}{}{}{}{}


\title{OBSERVATION OF SINGLE TOP QUARK PRODUCTION\\
       AT THE TEVATRON COLLIDER}

\author{\footnotesize A.P. HEINSON}

\address{Department of Physics and Astronomy, University of California,
Riverside\\ Riverside, California 92521-0413, USA\\
ann.heinson@ucr.edu}

\maketitle

\pub{\vspace{-0.02in}Received: February 8, 2010\vspace{-0.02in}}{}

\begin{abstract}
On March 4, 2009, the {\dzero} and CDF collaborations at Fermilab's
Tevatron Collider submitted papers to Physical Review Letters
announcing observation of single top quark
production.\cite{singletopdzeroobs}$^,$\cite{singletopcdfobs} This
review paper describes the successful searches carried out
independently by the two collaborations, allowing the reader to see
the similarities and differences that led to the simultaneous
discoveries. Both collaborations measured a cross section
$\sigma({\ppbar}{\rar}tb+X,tqb+X)$ consistent with the standard model
prediction at 5.0 standard deviation significance, and set a lower
limit on the quark mixing matrix element $|V_{tb}|$ without assuming
matrix unitarity with three quark generations.

\keywords{Single top quarks; electroweak production;
Tevatron collider.}
\end{abstract}

\ccode{PACS Nos.: 14.65.Ha, 12.15.Hh, 12.15.Ji, 13.85.Qk}


\section{Introduction to the Top Quark}

The top quark, an up-type quark, and the bottom quark, a down-type
quark, together form the third generation of quarks. Both top ($t$)
and bottom ($b$) quarks have spin~$1/2$. The top quark has electric
charge~$+2e/3$ and a mass of $173.1 \pm 1.3$~GeV.\cite{topmass} The
bottom quark has charge $-1e/3$ and mass
$4.20^{+0.17}_{-0.07}$~GeV.\cite{bottommass} All other quarks ($u$,
$d$, $c$, $s$) are nearly massless in comparison. The top quark has a
lifetime\cite{toplifetime} of $0.5 \times 10^{-24}$~s that is much
smaller than the strong interaction timescale, and is thereby unique
in the quark family, decaying before it can form a bound state with
another quark.\cite{tophadrons} Thus, the kinematics of the particles
from the top quark decay contain information about the bare top quark
itself. The Cabibbo-Kobayashi-Maskawa (CKM) matrix ``$V$'' describes
quark mixing.\cite{CKMmatrix} When there are exactly three quark
generations, the matrix is unitary, and a global fit to all available
precision data constrains the element $|V_{tb}|$ to be very close
to~one.\cite{Vtb1} Therefore, in the standard model (SM) the top quark
decays almost every time to a $W$~boson and a $b$~quark. The tiny SM
values for $|V_{td}|$ and $|V_{ts}|$ indicate that decays to $Wd$ and
$Ws$ are extremely rare.\cite{Vtb1} If there were a fourth quark
generation ($t^{\prime}$, $b^{\prime}$), then decays to light quarks
could occur more often since $|V_{tb}|$ would no longer be constrained
to have a value near~one.

\subsection{Production of Top Quarks at the Tevatron}

\begin{figure}[!h!tbp]
\begin{center}
\epsfig{file=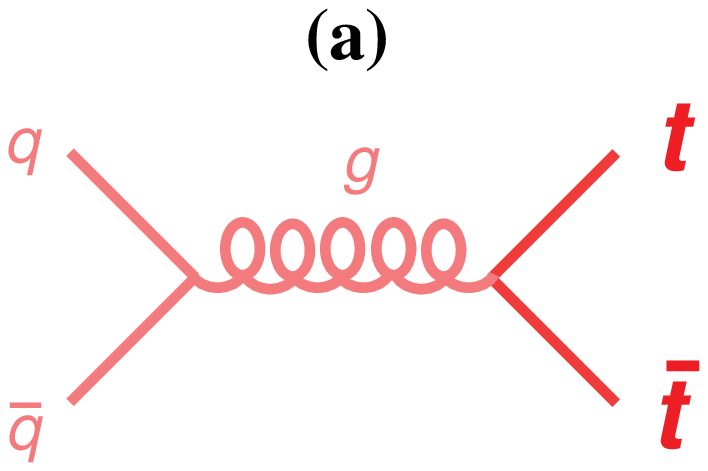,width=1.3in}
\hspace{0.1in}
\epsfig{file=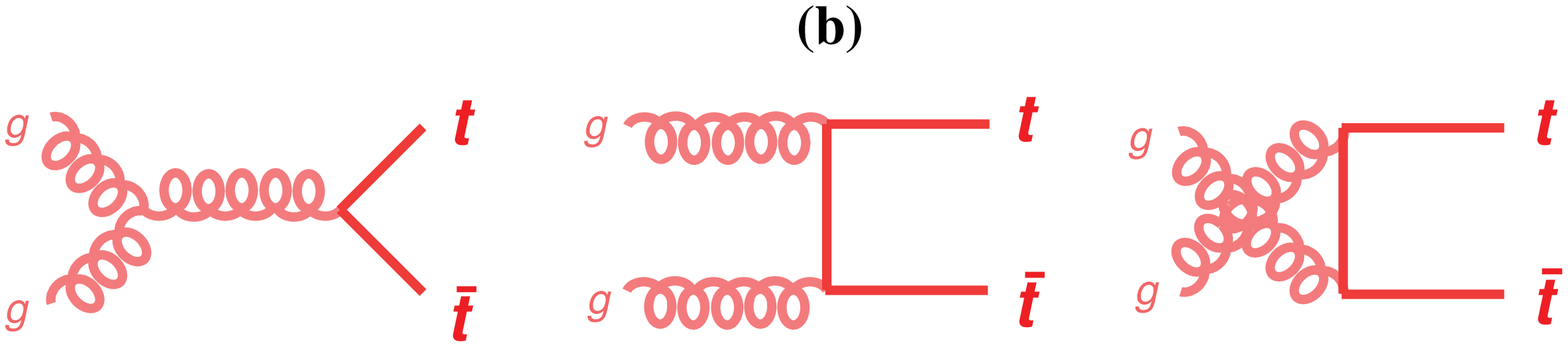,width=3.4in}
\\
\epsfig{file=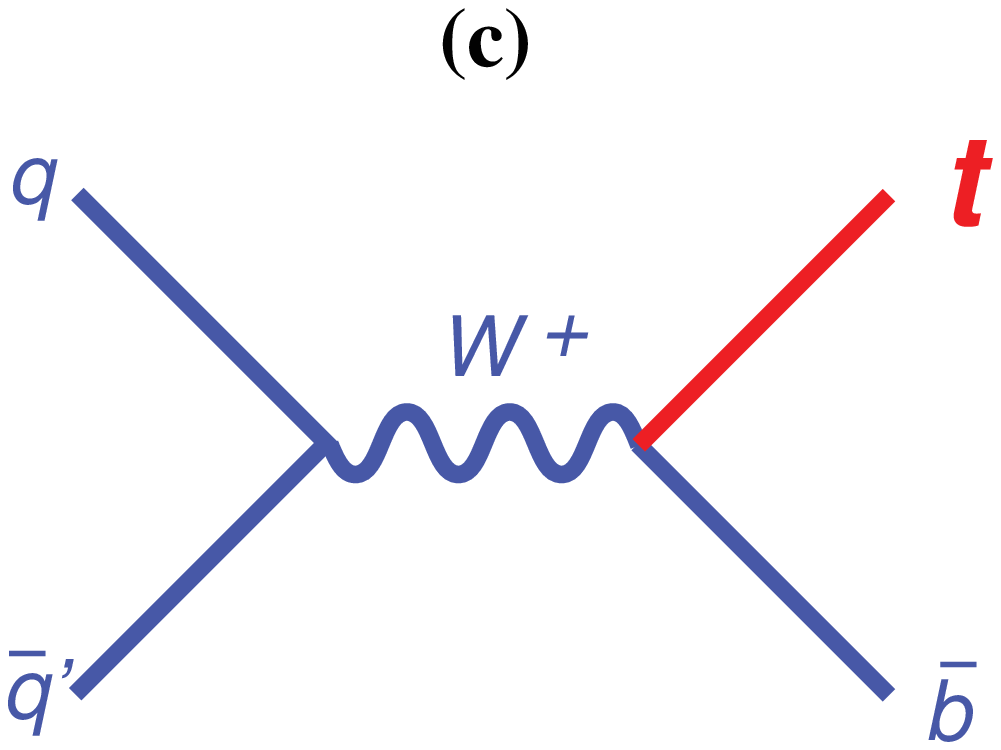,width=1.5in}
\hspace{0.1in}
\epsfig{file=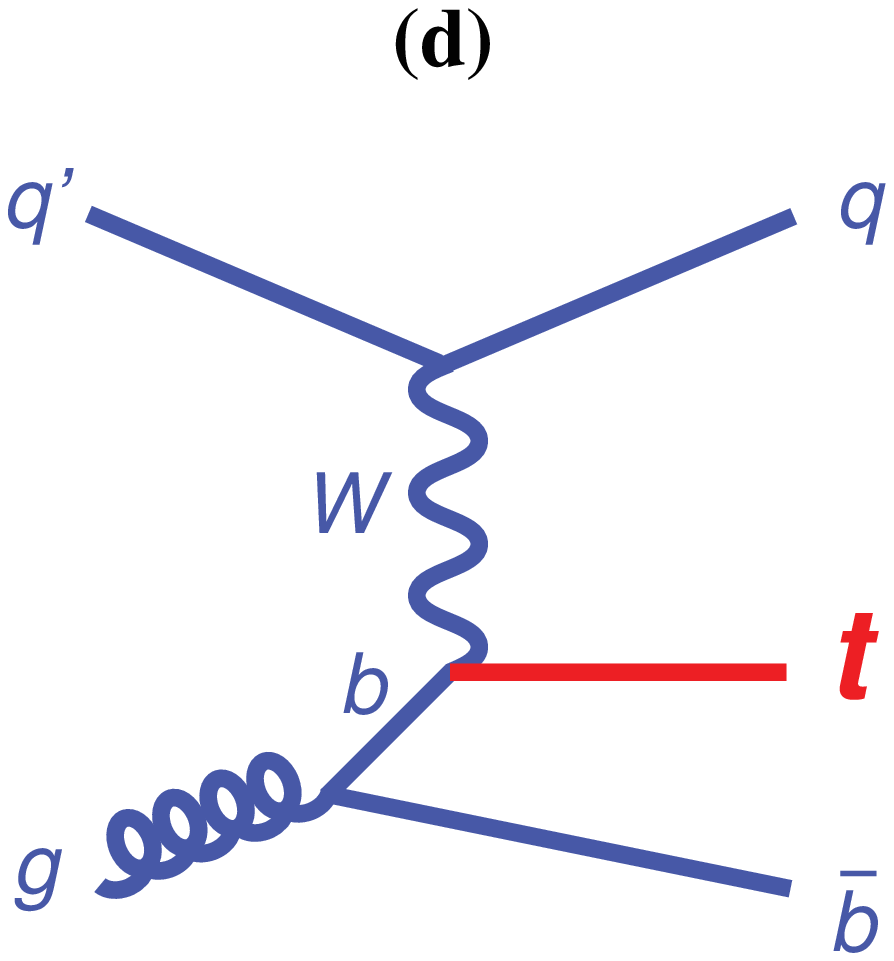,width=1.5in}
\hspace{0.1in}
\epsfig{file=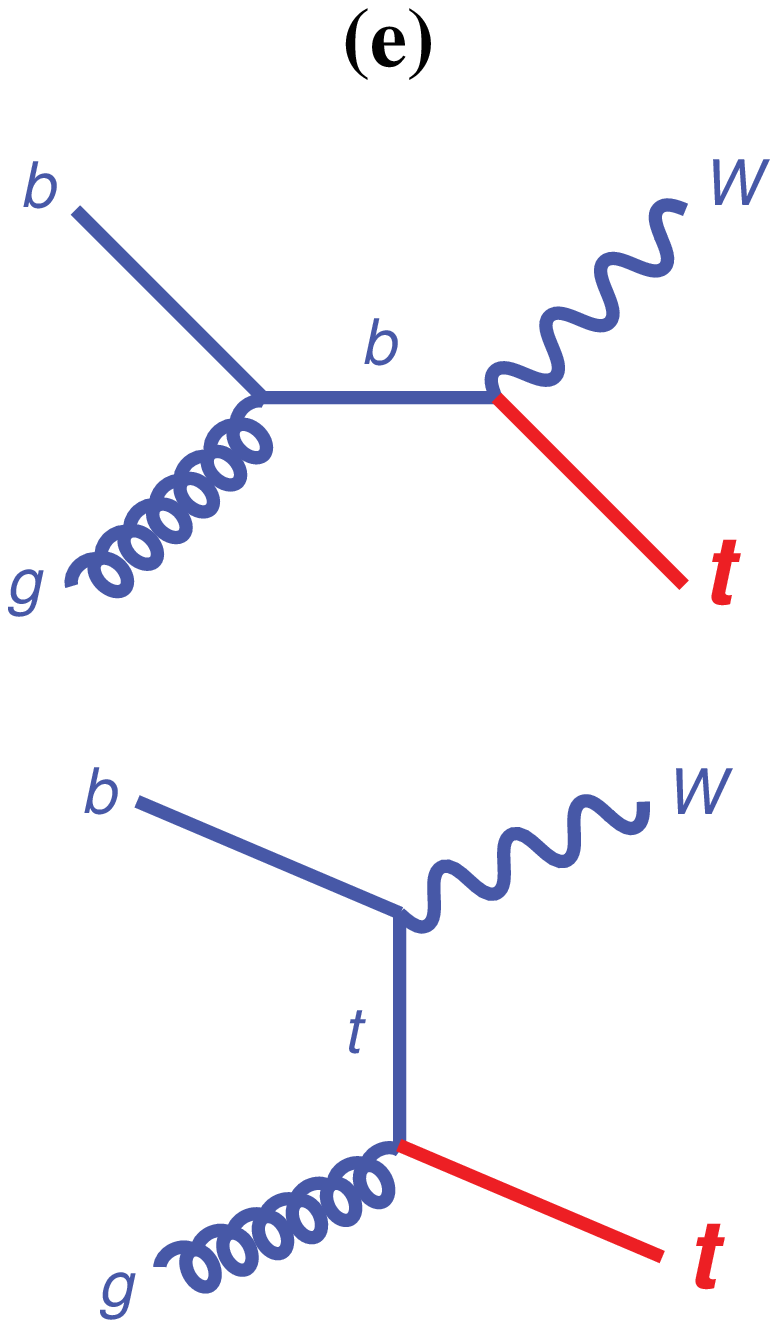,width=1.5in}
\end{center}
\vspace{-0.2in}
\caption[feynmantop]{Representative leading order Feynman diagrams for
strong production of top quark pairs from (a) quarks and (b) gluons,
and for electroweak production of single top quarks from (c) s-channel
``$tb$'' production, (d) t-channel ``$tqb$'' production, and (e)
``$tW$'' production. Process (a) produces 85\% of the {\ttbar} rate at
the Tevatron. Process (e) is not observed there as the cross section
is too low. The notation ``$tb$'' refers to the $t\bar{b}$ and charge
conjugate $\bar{t}b$ processes together; ``$tqb$'' refers to
$tq\bar{b}$ and $\bar{t}\bar{q}b$, and ``$tW$'' refers to $tW^-$ and
$\bar{t}W^+$. \label{feynman-top}}
\end{figure}

Quarks are sensitive to both the strong and electroweak forces. The
strong force is far more powerful than the electroweak force, and thus
top quarks are produced most often at hadron colliders via the decay
of a highly energetic virtual gluon to a top quark and a top antiquark
($\bar{t}$). The rate for this process is about 7~pb at the
Tevatron,\cite{ttbarxsec2008a}$^-$\cite{ttbarxsec2008d} where it was
first observed by the CDF and {\dzero} collaborations in
1995.\cite{topdiscoverycdf}$^,$\cite{topdiscoverydzero} Top quarks can
also be produced without their antiparticle partner via the
electroweak
interaction.\cite{singletopwillenbrock}$^-$\cite{singletopcampbell3}
In this case, a t-channel virtual $W$~boson and a highly energetic
bottom quark combine and produce a top quark, or a far off-shell
s-channel $W$~boson decays to produce a top quark and a bottom
antiquark. A third process is predicted to exist that occurs via both
the s-channel and t-channel, when a top quark is produced together
with a $W$
boson.\cite{singletopheinson}$^,$\cite{singletoptait}$^,$\cite{singletopkidonakis}
Charge-conjugate processes that produce top antiquarks are expected
via the same mechanisms. Contrary to expectations based on the
relative feebleness of the electroweak force, the rates for single top
quark production are calculated to be quite high, at about 2~pb for
the t-channel process and 1~pb for the s-channel
process.\cite{singletopharris}$^,$\cite{singletopkidonakis} This is
because higher-order corrections to the tree-level calculations for
the t-channel process are large. (The rate for $tW$ production is
predicted\cite{singletopkidonakis} to be about 0.3~pb and this process
is not seen at the Tevatron.) Therefore, one might expect searches of
the Tevatron data at the {\dzero} and CDF experiments to observe
s-channel and t-channel single top quark production rather easily,
given that the current datasets are 50 times larger than those used to
discover the top quark in 1995 using the pair production mode. The
reason that it has been very difficult to observe single top quark
production is not because the signal rate is too low, but because the
background processes are over 30 times larger than for top quark pair
events. The main leading order Feynman diagrams for strong and
electroweak production of top quarks at the Tevatron are shown in
Fig.~\ref{feynman-top}.

\section{The Search for Single Top Quarks at the Tevatron}

\subsection{The Tevatron Collider and the {\dzero} and CDF Experiments}

Fermi National Accelerator Laboratory is the home of the Tevatron
Collider, a 6.3~km circumference proton-antiproton ({\ppbar})
accelerator with superconducting magnets. This machine began operating
in collider mode (versus earlier provision of fixed target beams) in
1985 with only the Collider Detector at Fermilab (CDF) in operation to
record the collisions. The {\dzero} experiment started operation in
1992, at which time the collision energy was 1.8~TeV. This was raised
in 2001 to 1.96~TeV.

The two large multipurpose detectors have similar
structures:\cite{dzerorun1}$^,$\cite{dzerorun2} they consist of
concentric layers of detectors tightly packed around the beampipe at a
Tevatron collision region to a height of 9~m, with each detector layer
having a different purpose. The inner layers are composed of silicon
microstrip detectors, which provide the three-dimensional positions
where charged particles pass through. {\dzero} has an outer tracking
detector of scintillating fibers and CDF has an outer gaseous wire
drift chamber. Each of these tracking systems is encased in a solenoid
magnet with field-lines parallel to the beampipe; {\dzero}'s field
strength is 2.0~Tesla and CDF's is 1.4~T. The magnetic fields curve
the tracks of charged particles, which enables their transverse
momentum to be measured. {\dzero}'s magnet has a much smaller radius
(60~cm) compared to CDF's (150~cm) since it was retrofitted inside the
calorimeter in 2001. It therefore does not allow as much space for the
tracking detectors, resulting in far fewer hits per track, which makes
pattern recognition difficult with resulting lower track
reconstruction efficiency and higher fake track rates. The momentum
resolution in {\dzero} is poorer because of the shorter track length.

Outside the central magnets, each detector has layers of calorimetry
used to measure the energy of particles and to distinguish between
electromagnetic particles (electron and photons, with or without
matching central tracks), and jets (from quarks and
gluons). {\dzero}'s liquid-argon/uranium calorimeter is more hermetic
and covers a larger angular region (pseudorapidity $|\eta| < 4.2$
versus CDF's $|\eta|<3.6$, where $\eta = -\ln[\tan(\theta/2)]$ and
$\theta$ is the polar angle), giving better acceptance for forward
jets and missing transverse energy resolution.

Outside the calorimeter, {\dzero} has a muon spectrometer with up to
four layers of tracking detectors and a magnetic field strength of
1.9~T. Only muons (and invisible neutrinos) pass right through the
calorimeter and their position and momentum is remeasured here, with
wide pseudorapidity coverage to $|\eta| < 2.0$. CDF also has several
layers of muon detectors outside its calorimeters with coverage to
$|\eta| < 1.6$.

Each detector has a sophisticated multilevel trigger system used to
select interesting events from the {\ppbar} collisions, which occur
every 396~ns.

\vspace{-0.1in}

\subsection{Search History}

Preparation for a search for single top quark production began at the
{\dzero} experiment in 1994\cite{iowaheinson} (five months before the
first observation of top quarks in pair production mode). {\dzero}
published the results of a search using simple kinematic event
selection to set upper limits on the cross sections in
2000,\cite{singletopdzero1}$^-$\cite{qfthepheinson} with a follow-up
analysis making first use of a multivariate analysis technique, neural
networks, to separate signal from background in
2001.\cite{singletopdzero2} These analyses used 90~pb$^{-1}$ of data
at $\sqrt{s}=1.8$~TeV from Run~I at the Tevatron (1992--1996), when
{\dzero} did not have a silicon vertex detector or central magnetic
field and $b$~jets were identified via the presence of muons in jets
from the $b$~decay. The CDF collaboration also searched Tevatron Run~I
data for single top quark production; they published upper limits on
the cross sections from a cut-based selection in
2002\cite{singletopcdf1} and a follow-up analysis of the same dataset
using neural networks in 2004.\cite{singletopcdf2} CDF's analyses had
the advantage of being able to use secondary vertex $b$-jet
identification using the Silicon Vertex Detector.\cite{siliconcdf1}
The limits on the cross sections for s-channel and t-channel
production were about 10--20 times greater than the predicted values.

The Tevatron collision energy was increased to 1.96~TeV in 2001, and
the beam intensity was improved by a factor of about 15 over the
course of the run (2002--present). The {\dzero} and CDF detectors were
significantly upgraded, with the addition amongst other things, of the
central solenoid magnet to {\dzero} and very large silicon tracking
systems at both experiments.\cite{silicondzero}$^,$\cite{siliconcdf2}
CDF analyzed 160~pb$^{-1}$ of Run~II data using a cut-based selection
and a maximum-likelihood fit to the variable ``lepton charge $\times$
untagged jet pseudorapidity'' and set 95\% confidence level (CL) upper
limits of 13.6 pb on s-channel production and 10.1~pb on t-channel
production in 2005.\cite{singletopcdf3} {\dzero} analyzed
230~pb$^{-1}$ of data using neural networks (NN) for signal-background
separation and a Bayesian binned likelihood calculation using the NN
output distributions, and set 95\% CL upper limits of 6.4~pb in the
s-channel and 5.0~pb in the t-channel in
2005.\cite{singletopdzero3}$^,$\cite{singletopdzero4}

The next step in the search led to a major improvement. The {\dzero}
collaboration increased its dataset by a factor of four to
0.9~fb$^{-1}$, switched the search to $tb$+$tqb$ combined (assuming
the SM ratio of the two parts), loosened the selection cuts and used
an improved $b$-jet identification algorithm to increase the signal
acceptance by 13\% over that obtained in the earlier analysis, and
applied three multivariate methods to separate signal from background
to reach 3.4~standard deviation ($\sigma$) significance for a single
top quark signal. The measured cross section for $tb$+$tqb$ production
combined was $4.9 \pm 1.4$~pb. The measurement significance represents
a probability of 0.035\% for the background to have fluctuated up and
given a false measurement of signal with a cross section of at least
4.9~pb. A significance greater than 3$\sigma$ is considered in the
high energy physics community not to be sufficient for a claim of
discovery or first observation (which is set at 5$\sigma$), but is
high enough to indicate that ``evidence'' for the process in question
has been seen; it is a very exciting threshold to reach. The result
was published in 2007 and has received well over 100 citations to
date.\cite{singletopdzero5} Small improvements were made to the
analysis and a slightly more significant result (3.6$\sigma$) was
published in a long paper in
2008.\cite{singletopdzero6}$^,$\cite{hcpheinson} The CDF collaboration
performed a similar analysis on 2.2~fb$^{-1}$ of data and reached a
significance for single top quark signal of 3.7$\sigma$, published in
2008.\cite{singletopcdf4} They measured the cross section for
$tb$+$tqb$ production to be $2.2 \pm 0.7$~pb.


\section{Measurement Overview}

After the ``evidence'' papers, the {\dzero} and CDF collaborations
each worked to improve their analysis methods and apply them to larger
datasets. Both collaborations select events with one isolated high
transverse momentum ($p_T$) lepton (electron or muon) and large
missing transverse energy ({\met}), indicative of a leptonic $W$-boson
decay, together with two, three, or four jets. One or two of the jets
is identified as originating from a $b$~quark, which could be from the
top quark decay or produced together with it.

The CDF collaboration has an additional independent search
channel\cite{singletopcdfmet} that requires no identified charged
lepton, which picks up events lost to electron or muon identification
inefficiencies, and some $\tau$+jets events where the $\tau$ decayed
hadronically (but there was no explicit $\tau$ reconstruction). This
is the first time that the {\met}+jets channel has been used in a
single top quark measurement.

Both collaborations include signal and background events in their
lepton+jets channels with $t{\rar}Wb$, $W{\rar}\tau\nu_{\tau}$, and
$\tau{\rar}e\nu_{e}\nu_{\tau}$ or $\tau{\rar}\mu\nu_{\mu}\nu_{\tau}$.
Neither includes events with $\tau{\rar}{\rm hadrons}$ in the signal
acceptance since hadronic $\tau$ reconstruction is difficult. A new
search based just on this decay channel has recently been completed by
{\dzero}.\cite{singletopdzerotau}

\begin{figure}[!h!tbp]
\begin{center}
\epsfig{file=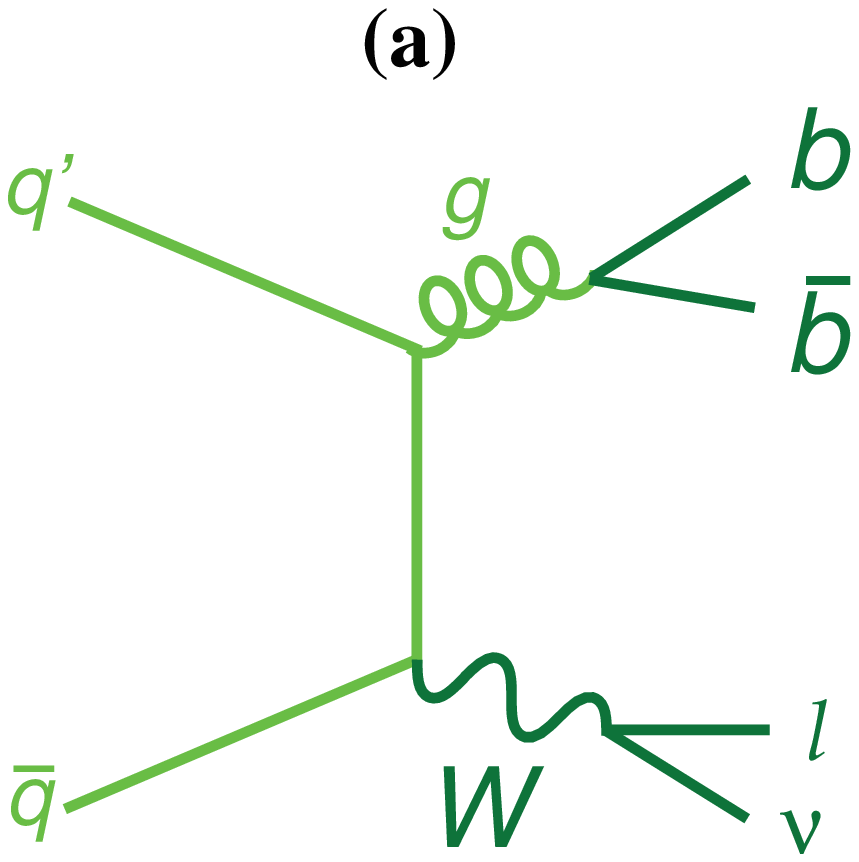,width=1.15in}
\hspace{0.02in}
\epsfig{file=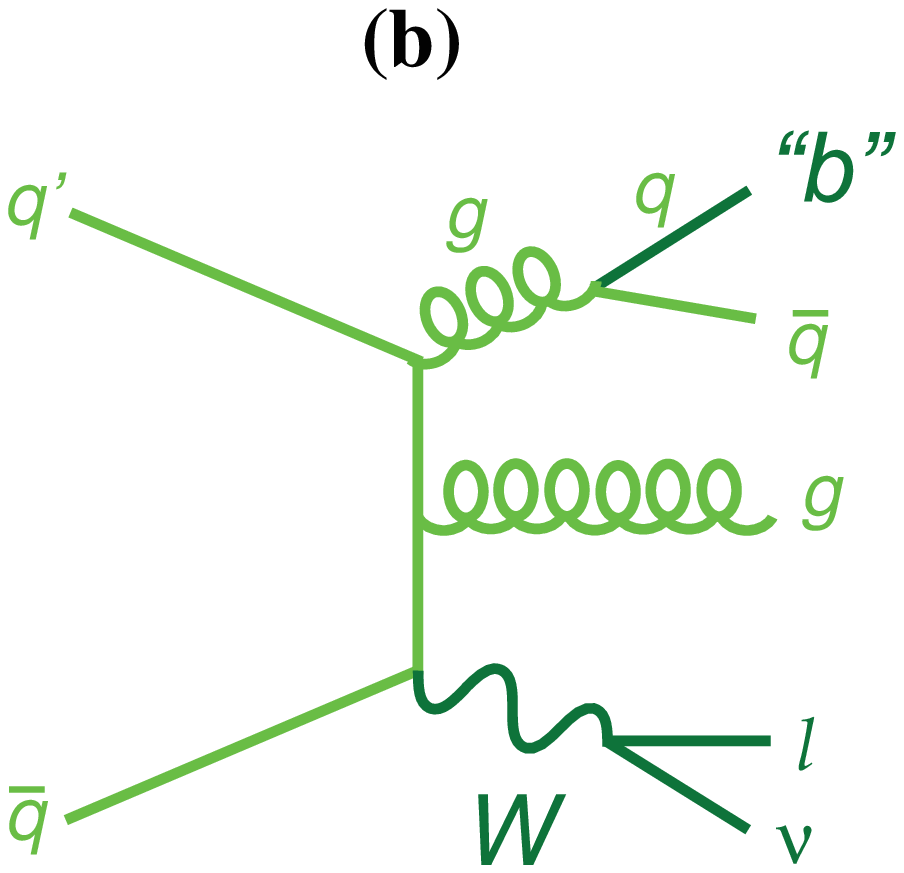,width=1.15in}
\hspace{0.02in}
\epsfig{file=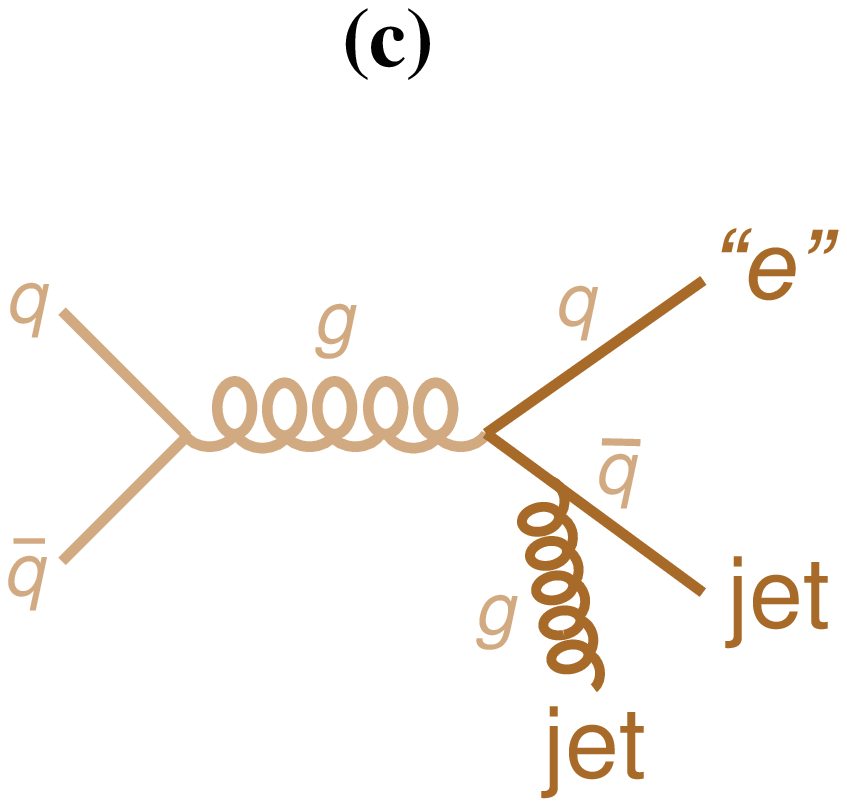,width=1.15in}
\hspace{0.02in}
\epsfig{file=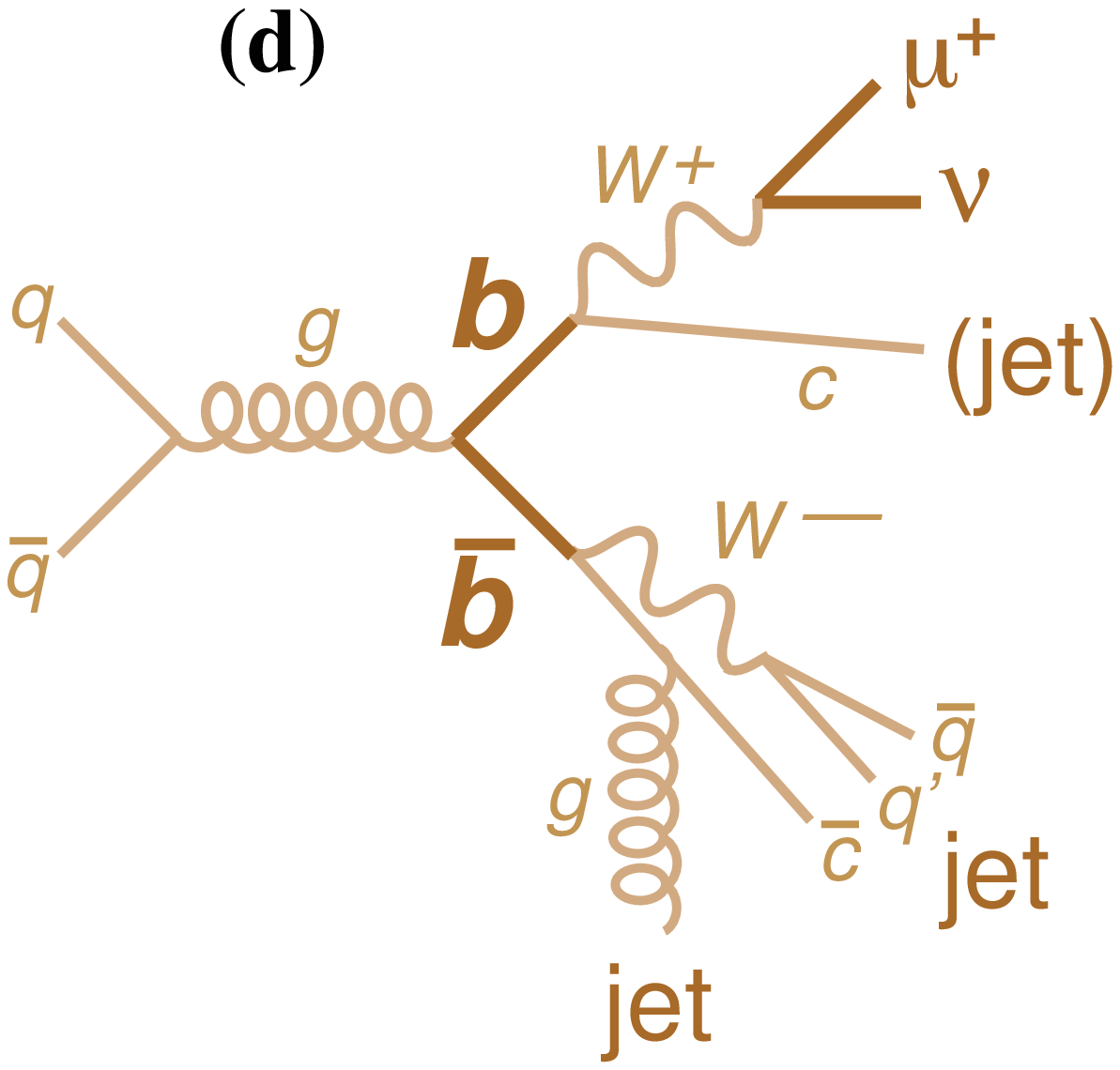,width=1.15in}
\end{center}
\vspace{-0.1in}
\caption[feynmanbkgd]{Representative leading order Feynman diagrams for
the background processes: (a)~$W$+jets with real $b$~jets,
(b)~$W$+jets with a light jet mistagged as a $b$~jet, (c)~multijets
with a jet misidentified as an electron, and (d)~{\bbbar} multijets
with a nonisolated muon (from a $b$~decay) misidentified as an isolated
one (from a $W$~decay). \label{feynman-backgrounds}}
\end{figure}

After event selection, the signal-to-background ratio is approximately
1:20. The backgrounds are mostly $W$+jets events (especially at low
jet multiplicity), followed by {\ttbar} pairs (especially at high jet
multiplicity), with small contributions from $Z$+jets, dibosons ($WW$,
$WZ$, $ZZ$), and multijets. Top pairs look like signal when one
$W$~boson decays leptonically ($e\nu$ or $\mu\nu$) and the other
decays hadronically ($u\bar{d}$, $c\bar{s}$, etc.) producing
lepton+jets events, and also when both $W$'s decay leptonically and
event reconstruction fails to identify one of the leptons. $Z$+jets
events and some diboson processes also mimic single top quark signals
when the $Z$~boson decays to a pair of leptons ($e^+e^-$ or
$\mu^+\mu^-$) and one of the leptons is lost, generating fake
{\met}. Multijet events look like signal in the electron channel when
a jet is misidentified as an electron and a jet's energy is
mismeasured, creating false {\met}. In the muon channel, the multijet
background comes mostly from {\bbbar} events where one of the $b$'s
decays to a muon that travels wide of its jet or the jet is not
reconstructed (its energy is too low maybe). Example Feynman diagrams
for the $W$+jets and multijets processes are shown in
Fig.~\ref{feynman-backgrounds}. After event selection, {\dzero} has
4,519 lepton+jets events and CDF has 3,315 lepton+jets events and
1,411 {\met}+jets events.

\section{Data Samples}

For the observation analysis, {\dzero} uses 2.3~fb$^{-1}$ of Run~II
data, collected from August 2002 until August 2007. The dataset is
split in two parts (``Run~IIa'' and ``Run~IIb'') to denote a
significant upgrade to the detector, when a new layer of silicon
microstrip detectors was added around the
beampipe.\cite{silicondzerol0} This improved the tracking and
\mbox{$b$-tagging} efficiencies. The Run~IIb half dataset has higher
instantaneous luminosity than the Run~IIa half, which lowers the
primary vertex identification efficiency and increases track
multiplicities in the events. The CDF collaboration uses a
3.2~fb$^{-1}$ dataset for the lepton+jets analysis, which is not split
like {\dzero}'s. The {\met}+jets analysis uses 2.1~fb$^{-1}$ of data.

{\dzero} selects data that pass any reasonable trigger. This
requirement is relaxed from earlier analyses where only lepton+jets
triggers were used. The change increases the signal acceptance by 16\%
in the electron channel and 20\% in the muon channel. It also
increases the trigger efficiency to $\approx100\%$, meaning that no
correction functions are needed to model trigger turn-on curves for
Monte Carlo events. CDF's triggers include a high-$p_T$ electron
trigger, a high-$p_T$ muon trigger, and one that requires high {\met}
and either an energetic electromagnetic cluster or two jets.


\section{Signal and Background Simulation}

\subsection{Single Top Quark Signal Models}

The single top quark signal is modeled to reproduce next-to-leading
order (NLO) kinematics\cite{singletopsullivan} using modified leading
order (LO) generators. {\dzero} uses {\sc
singletop},\cite{mcsingletop} a version of {\sc
comphep}\cite{mccomphep} adapted by its authors for {\dzero}, and CDF
uses {\sc madevent}\cite{mcmadevent} based on {\sc
madgraph}\cite{mcmadgraph} with their own modifications. In fact,
s-channel simulation at LO reproduces NLO kinematics without
changes,\cite{singletopsullivan} and it is only the t-channel that
needs such attention. The transverse momentum distribution of the
bottom antiquark in the $2{\rar}2$ process $q^{\prime}b{\rar}tq$ (from
{\sc singletop} or {\sc madevent}) after back-propagation of the
initial-state $b$ to $g{\rar}b\bar{b}$ (from {\sc
pythia}\cite{mcpythia}) is matched to that of the $\bar{b}$ in the
$2{\rar}3$ process $q^{\prime}g{\rar}tq\bar{b}$ (from {\sc singletop}
or {\sc madevent}). Simulated events from the $2{\rar}2$ calculations
are kept if $p_T(\bar{b}) \le 10$~GeV ({\dzero}) or $\le 20$~GeV (CDF)
and ones from the $2{\rar}3$ process are used if $p_T(\bar{b}) >
10$~GeV ({\dzero}) or $> 20$~GeV (CDF). The $2{\rar}2$ process is
scaled by a $K$ factor to make the rates at the cut-off point match:
$K = 1.21$ for {\dzero}.\cite{mcsingletop} There is another t-channel
subprocess where a gluon produces a {\ttbar} pair and the $\bar{t}$
combines with a radiated $W$~boson to produce a
$\bar{b}$.\cite{singletopheinson} This subprocess has a cross section
only a few percent of the $tqb$ $g{\rar}{\bbbar}$ subprocess, with a
large negative interference between the two subprocesses. Both
{\dzero}'s {\sc singletop} and CDF's {\sc madevent} models include the
$g{\rar}{\ttbar}$ subprocess and the interference. The models also
both have finite widths for the top quark ($\approx 1.5$~GeV) and
$W$~boson ($\approx 2.0$~GeV). In all signal (and {\ttbar} background)
models, the top quarks and their daughter $W$~bosons are decayed at
the time of production, before later processing with {\sc pythia}, so
that all spin properties of the top quarks are preserved in the
angular correlations of the final decay products. {\dzero} uses a top
quark mass of 170~GeV for signal simulation, CDF uses 175~GeV. Each
value was chosen at a time when it was close to the world average
value, which shifts slightly once or twice a year as the measurement
is improved. This difference does not have a significant effect on the
final results.

For modeling the parton kinematics in the protons and antiprotons,
{\dzero} uses the CTEQ6M next-to-leading-order parton distribution
functions (PDF).\cite{cteq6} CDF uses the CTEQ5L leading order
PDFs.\cite{cteq5} The $Q^2$ scale for the s-channel model is $M_{\rm
top}^2$ ({\dzero}) or $\hat{s}$ (CDF), and for the t-channel model,
$(M_{\rm top}/2)^2$ ({\dzero}) or $\hat{t}+M_{\rm top}^2$ (CDF) are
used. {\dzero}'s values in {\sc singletop} are chosen to make the LO
and NLO cross sections be the same\cite{singletopstelzer2} and CDF's
values in {\sc madevent} are chosen to closely match those used in the
NLO {\sc ztop} event generator.\cite{singletopsullivan} Both
collaborations use {\sc pythia} to add the underlying event from the
{\ppbar} interaction, the initial-state and final-state radiation, and
to hadronize and fragment the final state quarks and gluons into
jets. They also both use {\sc tauola} to decay tau
leptons.\cite{tauola} For $B$-hadron decay, {\dzero} uses {\sc evtgen}
from the BaBar experiment\cite{evtgen} and CDF uses {\sc qq} from the
CLEO experiment.\cite{qq} Events from multiple primary vertices are
overlaid onto the primary MC event with a Poisson multiplicity
distribution in order to simulate the high instantaneous
luminosity. {\dzero} uses zero-bias data events and CDF uses MC events
generated with {\sc pythia}. The mean number of {\ppbar} collisions
per bunch crossing for this dataset is two for Run~IIa and five for
Run~IIb.

\vspace{-0.1in}
\subsection{Background Models}

All background components except multijet events are simulated using
Monte Carlo models. Both collaborations use the {\sc alpgen} event
generator,\cite{mcalpgen} which has leading-log (LL) precision,
coupled to {\sc pythia} to simulate $W$+jets events, including full
modeling of events with massive $b$ and $c$ jets. The version of {\sc
alpgen} used includes parton-jet matching\cite{matching} to avoid
double-counting some regions of jet kinematics. The samples are
generated in the following sets (lp = light partons): $W$+0lp,
$W$+1lp, $W$+2lp, $W$+3lp, $W$+4lp, $W$+$\ge$5lp (this set includes
$W$+single massless charm); $W{\ccbar}$+0lp, $W{\ccbar}$+1lp,
$W{\ccbar}$+2lp, $W{\ccbar}$+$\ge$3lp; and $W{\bbbar}$+0lp,
$W{\bbbar}$+1lp, $W{\bbbar}$+2lp, $W{\bbbar}$+$\ge$3lp, which are
summed weighted by the {\sc alpgen} LL average cross section for each
subset and then split to obtain $W$+2jets, $W$+3jets, and $W$+4jets
event sets. {\dzero} uses this same version of {\sc alpgen} and
generation method to simulate {\ttbar} events; CDF uses {\sc pythia},
which only adds extra jets through initial-state and final-state
radiation (not from the hard scatter), but this is not critical since
they do not include events with four jets in their analyses. Smaller
backgrounds are modeled using {\sc alpgen} and {\sc pythia} ({\dzero})
and {\sc pythia} (CDF).

Some more details of the $W$+jets modeling are in order, since this
background is critical in the most important 2-jets analysis
channels. {\dzero} uses the CTEQ6L1 parton distribution functions and
CDF uses CTEQ5L. Both collaborations use scale $Q^2 = m_W^2 + \sum
m_T^2$ (recommended by the {\sc alpgen} authors), where $m_T$ is the
transverse mass defined as $m_T^2 = m^2({\rm parton}) + p_T^2({\rm
parton})$ and the sum $\sum m_T^2$ extends to all final state partons
(including the heavy quarks, excluding the $W$ decay products). For
$W{\bbbar}$ and $W{\ccbar}$ samples, $m({\rm parton}) = m_b$ or
$m_c$. For $W$+light jets samples, the jets are treated as massless
with $m({\rm parton}) = 0$~GeV.

The multijet background is modeled by {\dzero} using data with much
looser lepton selection than for signal selection. They select events
that pass all final cuts in the electron channel except that the
electromagnetic object fails the electron identification cuts,
including not requiring a track matched between the primary vertex and
energy cluster in the calorimeter. This is a very loose selection,
with a ten-fold increase in statistics compared to that used in the
earlier evidence analysis (when a matching track was required). The
reason for this change is to ensure sufficient statistics after
$b$~tagging to make a proper measurement of this background. In
{\dzero}'s muon multijet data, the muon is not required to be isolated
from a jet. In the previous analysis, a partial isolation requirement
was applied, and removing this criterion increased the muon multijet
background statistics by a factor of ten. In the electron channel, the
ratio of electron+jets and photon+jets events is used to determine a
reshaping weight as a function of the electron $p_T$ to make the
background model sample better match the actual multijet events
remaining after signal selection. The function boosts the fraction of
low-energy events. In the muon multijet background dataset, any jets
close to the muon are removed and the {\met} is recalculated in order
to make the jets reproduce those in the signal data. No kinematic
reshaping is needed. To obtain background samples that model the
multijets backgrounds, the samples as described (with an electron or
muon that fails final identification criteria) are scaled by functions
that represent the probability for a failing lepton to pass the
identification cuts.

CDF model the multijets background in the lepton+jets channels using a
data sample with {\met} below the signal selection threshold of
25~GeV, and project it into the high-{\met} signal region using a fit
to the shape of the {\met} distribution. They model the dominant
multijets background in the {\met}+jets channels using pretagged data
that pass all selection cuts together with a tag-rate matrix
calculated using an independent {\met}+jets dataset.

\subsection{Detector Simulation}

After the MC samples are generated, they are processed through code
that models the geometry and material of each subdetector
system,\cite{geant} and then through further code that generates
digitized signals modeled to resemble those from the
subdetectors. After this, the MC events look very like those from data
and both are processed through event reconstruction software to
identify the correct primary vertex, leptons, jets, and so on, ready
for further analysis.

\subsection{Background Normalization}

The {\ttbar}, $Z$+jets, and diboson backgrounds are normalized to
(N)NLO theory cross section values, with each collaboration using a
{\ttbar} cross section appropriate for the top quark mass chosen for
its analysis.

Before the $W$+jets backgrounds can be normalized, corrections are
applied to modify the leading log {\sc alpgen} fractions of heavy
flavor jets ($c$, ${\ccbar}$, and ${\bbbar}$) to account for missing
higher order contributions and make them match what is seen in data.
The {\dzero} collaboration scales the $Wjj$, $Wcj$, $W{\ccbar}$ and
$W{\bbbar}$ subprocesses to their NLO predictions (where $j$ = $u, d,
s, g$) using $K^{\prime} = \sigma_{\rm NLO}/\sigma_{\rm LL}$ and
$K^{\prime}_{\rm HF} = \sigma_{\rm NLO}^{\rm HF}/\sigma_{\rm NLO}$
factors. This is also done for the small $Z$+jets background. The
$K^{\prime}$ factor is 1.30. $K^{\prime}_{\rm HF} = 1.47$ for
$W{\ccbar}$ and $W{\bbbar}$, 1.67 for $Z{\ccbar}$, and 1.52 for
$Z{\bbbar}$. These factors come from calculations using the NLO MC
event generator {\sc mcfm}.\cite{mcmcfm} For $Wcj$, $K^{\prime}_{\rm
HF} = 1.38$, from a data measurement that agrees with NLO
theory.\cite{dzeroWcjWjj} The important $W{\bbbar}$ and $W{\ccbar}$
subprocesses are then checked against data after $b$~tagging and an
empirical correction of $0.95\pm0.13$ is applied to get good
data-background agreement. This factor accounts for contributions to
the heavy flavor rate from Feynman diagrams at higher order than NLO
not included in {\sc mcfm}. The uncertainty on the empirical
correction factor is the third largest component of the total
systematic uncertainty on the cross section measurement. It includes a
9\% statistical contribution from the variation of the correction when
measured in different analysis channels ($e$, $\mu$, 1-tag, 2-tags,
Run~IIa, Run~IIb), 8\% from the $Wcj$ $K^{\prime}_{\rm HF}$ factor
uncertainty (10\%), and 7\% from the uncertainty on the assumed single
top cross section (40\%, based on the difference between {\dzero} and
CDF's published evidence measurements). CDF compresses these three
steps into one, and, from a data-background comparison after
$b$~tagging in $W$+1jet events, applies a scale factor of $1.4\pm0.4$
to $Wcj$, $W{\ccbar}$ and $W{\bbbar}$ relative to the LL $Wjj$
process. Converting {\dzero}'s scale factors to allow a comparison
gives $1.47\times0.95 = 1.40$, so things are consistent.

The $W$+jets and multijets backgrounds are normalized to data before
$b$~tagging. {\dzero} normalizes the sum of the two backgrounds using
an iterative Kolmogorov-Smirnov procedure with the $p_T$(lepton),
{\met}, and $W$~boson transverse mass $M_T(W)$ variables. For the
multijets background, CDF uses a fit to the {\met} distribution at low
{\met} extrapolated to high {\met} and does not anticorrelate the two
components. After subtracting all other background components, they
normalize the $W$+jets background to the number of data events.

\subsection{Model Corrections}

Both collaborations need to correct the MC efficiency to reproduce the
efficiency of the detector, event reconstruction, and particle
identification. This is done for electrons, muons, and jets. All MC
events are reweighted to make the instantaneous luminosity
distribution (number of overlaid zero-bias events from multiple
{\ppbar} collisions) match that observed in the data. {\dzero} also
reweights the muon pseudorapidity $\eta$ distribution in $W$+jets
events to better model the efficiencies of the regions between the
central and forward muon systems.

For $W$+jets events, both collaborations find the pseudorapidity
distributions of the jets from the {\sc alpgen} simulation do not
match data well (there are presumed to be slightly different Feynman
diagrams in the calculation compared with e.g., the {\sc sherpa}
model\cite{mcsherpa} which has wider jet $\eta$ distributions). The
{\sc alpgen} distributions are too narrow, and empirical reweightings
are applied to these distributions ($\eta({\rm jet1})$, $\eta({\rm
jet2})$, $\Delta\phi({\rm jet1, jet2})$, and $\Delta\eta({\rm jet1,
jet2})$ for {\dzero}, similarly for CDF) to make the background model
match data before $b$~tagging. Since {\dzero}'s reweighting uses
binned functions derived in each analysis channel separately, it also
takes account of imperfections of the detector model in the
intercryostat regions.


\section{Event Reconstruction and Particle Identification}

\subsection{Primary Vertices}

There are several primary vertices in each event, on average, because
of the high collision rate leading to multiple interactions. They are
reconstructed at {\dzero} by first clustering tracks according to
their positions along the beamline, then the location and width of the
beam is measured and used to refit the tracks. Finally, each cluster
of tracks is associated with a vertex, and the one with the lowest
probability of coming from a zero-bias collision is chosen as the
primary vertex for that event.

\subsection{Electrons}

Electrons are defined as clusters of energy deposited in the
electromagnetic section of the calorimeter that are consistent in
shape and other properties with an electromagnetic shower. The cluster
must be isolated from other energy in the event and have a track that
points to it from the primary vertex.

\subsection{Muons}

Muons are identified by matching reconstructed tracks from the outer
muon system to ones from the inner tracking system. The match is made
spatially and (at {\dzero}) in transverse momentum and muon
charge. Muons must be isolated from nearby tracks and jets to show
they are from $W$ (or $Z$) boson decay and not from heavy flavor ($b$
or $c$) decay inside a jet.

\subsection{Jets}

Jets are reconstructed using energy deposited in the calorimeters.
{\dzero} applies the midpoint cone algorithm\cite{jetalgorithmdzero}
in $(y,\phi)$ space, where $y$ is the rapidity and $\phi$ is the
azimuthal angle, and the cone radius is 0.5. CDF uses a clustering
algorithm\cite{jetalgorithmcdf} in $(\eta,\phi)$ space with a cone
radius of 0.4. There are several requirements on where the energy is
deposited to reject noisy jets (whose energy would be
mismeasured). The energy of each jet is corrected if there is a muon
in the jet, to account for energy taken away by that muon and
associated (invisible) neutrino from a heavy quark decay. The jet's
energy is also corrected using the jet energy scale calibration to
ensure that the absolute value is correct. For most jets ($E_T$,
$\eta$), the uncertainty on the jet energy scale is between 1\% and
2\% for {\dzero}\cite{jesdzero} and it is 3\% for CDF.\cite{jescdf}

\subsection{Missing Transverse Energy}

The missing transverse energy is computed by adding up vectorially the
transverse energies in all cells of the electromagnetic and fine
(inner) hadronic calorimeters (for {\dzero}). Cells in the coarse
(outer) hadronic calorimeter are only added if they form part of a
good jet. This quantity is corrected for all the energy corrections
applied to other objects in the event and for the momentum of isolated
muons. CDF's computation of {\met} is similar.


\section{Event Selection}

The analyses start out with very large numbers of events in data and
MC signal and background samples. For example, {\dzero} uses data
skims with one electron or one muon in them, which contain 1.2~billion
events, and 85~million MC events. From these samples, the analyses
first select events that look like signal and reject events that do
not. That is, each collaboration devises selection cuts designed to
keep as many MC signal events as possible while rejecting as much
background as they can. The {\dzero} collaboration chooses to maximize
signal acceptance while allowing for a slightly worse
signal-to-background ratio, whereas the CDF collaboration chooses
tighter selection cuts that produce a lower signal acceptance but
better signal-to-background ratio. Thus, although {\dzero} starts the
analysis with about 30\% less integrated luminosity to analyze than
CDF, they end up with more expected signal events in the lepton+jets
channel, and a similar number in total when considering also the
{\met}+jets channel after all selections are applied. {\dzero} pursues
this strategy because their studies show that the overall sensitivity
of the analysis is proportional to the signal acceptance.

\subsection{Kinematic Cuts}

The kinematic cuts used in the analyses are shown in
Table~\ref{selection-cuts}. For simplicity, only the cuts in the
channels with exactly two jets are shown, since these channels
contribute most to the analysis sensitivity. For the lepton+jets
analyses, both collaborations also use events with three jets, and
{\dzero} also uses events with four jets. These channels have slightly
harder cuts for electron $p_T$, {\met}, and total transverse energy
$H_T$ than those shown in the table, to reject the higher multijets
background.

\begin{table}[!h!tbp]
\tbl{Kinematic selection cuts used in the 2-jets analysis
channels to identify events that look like single top quark signal
and reject backgrounds. \label{selection-cuts}}
{\begin{tabular}{@{}llll@{}}
\toprule
       & {\dzero}'s Selection & \multicolumn{2}{c}{CDF's Selection} \\
       & Lepton+2Jets         & Lepton+2Jets    & {\met}+2Jets       \\
\colrule
Electron~~~ & $p_T > 15$~GeV    & $p_T > 20$~GeV~~~~~~~~~ & ~~~~~~---\\
            & $|\eta| < 1.1$    & $|\eta| < 1.6$    & ~~~~~~--- \\
Muon        & $p_T > 15$~GeV    & $p_T > 20$~GeV    & ~~~~~~--- \\
            & $|\eta| < 2.0$    & $|\eta| < 1.6$    & ~~~~~~--- \\
Neutrino    & ${\met} > 20$~GeV & ${\met} > 25$~GeV & ${\met}>50$~GeV\\
Jet1        & $p_T > 25$~GeV    & $p_T > 20$~GeV    & $p_T > 35$~GeV\\
            & $|\eta| < 3.4$    & $|\eta| < 2.8$    & $|\eta| < 0.9$\\
Jet2        & $p_T > 15$~GeV    & $p_T > 20$~GeV    & $p_T > 25$~GeV\\
            & $|\eta| < 3.4$    & $|\eta| < 2.8$    & $|\eta| < 2.8$\\
Total $E_T$ & $H_T({\rm jets},e,{\met})  > 120$~GeV~~~ & ~~~~~~--- & ~~~~~~--- \\
            & $H_T({\rm jets},\mu,{\met})> 110$~GeV & ~~~~~~--- & ~~~~~~--- \\
\botrule
\end{tabular}}
\end{table}

\begin{figure}[!h!tbp]
\begin{center}
\epsfig{file=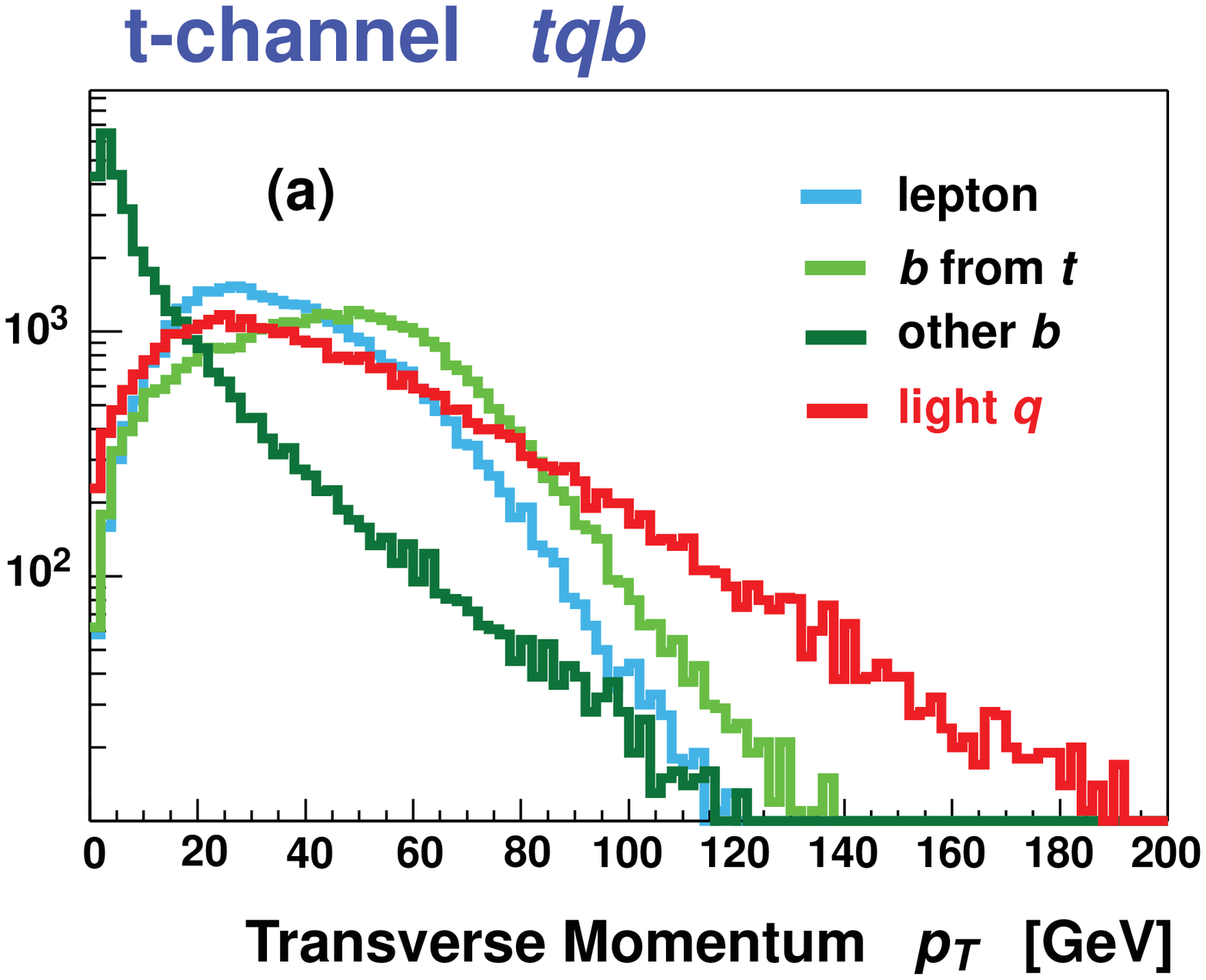,width=2.38in}
\hspace{0.1in}
\epsfig{file=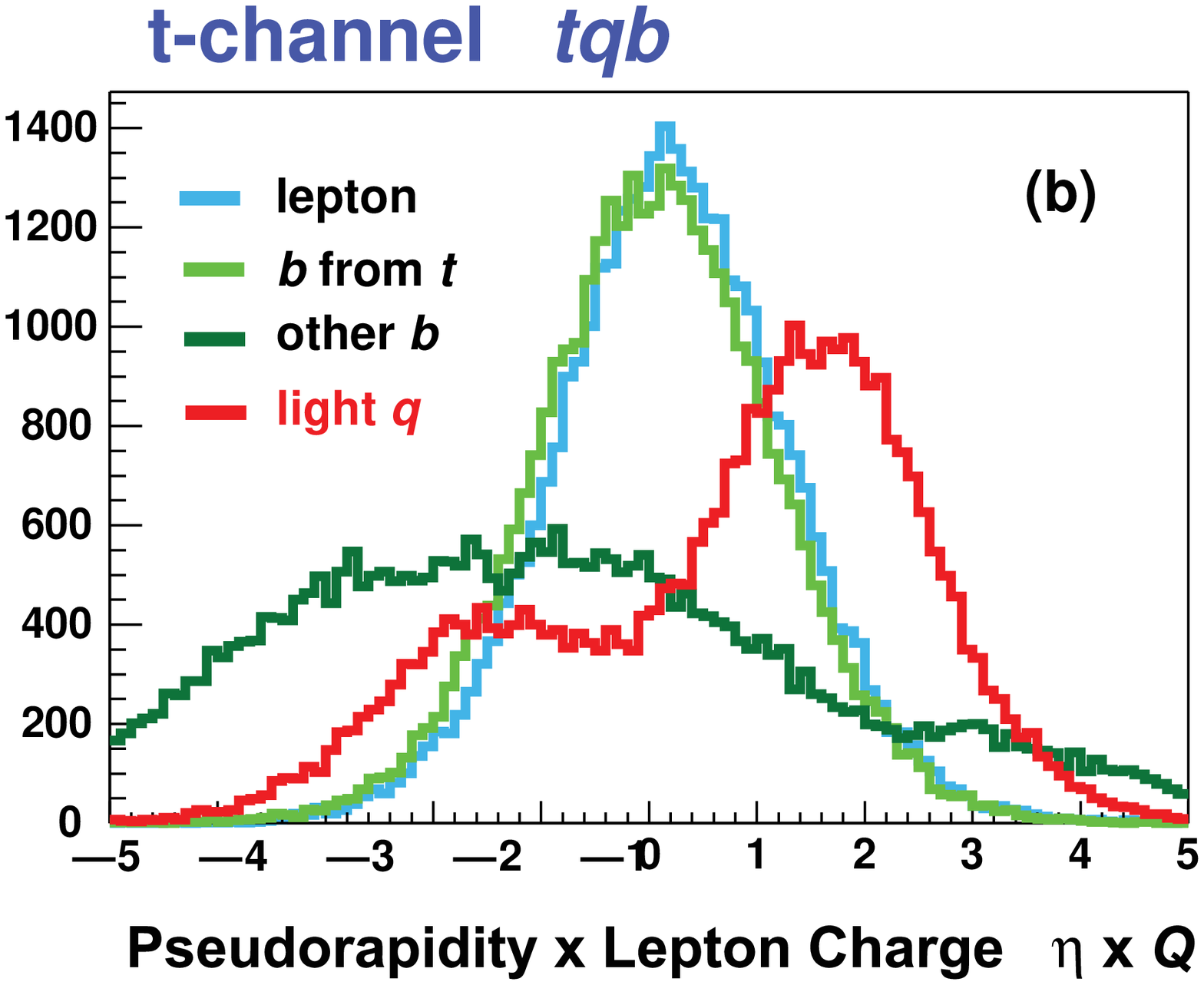,width=2.38in}
\end{center}
\vspace{-0.1in}
\caption[partons]{Distributions of (a)~the transverse momentum and
(b)~the signed pseudorapidity of partons in t-channel single top quark
events, from the {\sc comphep-singletop} simulation.
\label{parton-kinematics}}
\end{figure}

Motivation for {\dzero}'s choice of lower transverse energy thresholds
and wider jet pseudorapidity distributions than used in, for example,
a top pairs measurement can be seen in Fig.~\ref{parton-kinematics}
for the t-channel single top quark process. The light quark that
radiates the $W$~boson has a very wide $\eta$ distribution in both the
forward and backward directions (shown by the red histograms in the
plots). This is a very strong signature for single top quark
production that will be used as a powerful variable to separate signal
from background. The soft $\bar{b}$ produced from the gluon splitting
has an even wider $\eta$ distribution (dark green histograms) and low
$p_T$, and finding this jet increases the double-$b$-tagged signal
acceptance.

There are additional selection cuts not shown in
Table~\ref{selection-cuts}. In the lepton+jets channels, events are
rejected if there is a second isolated lepton, which rejects dilepton
decays of {\ttbar}, $Z$+jets, and diboson events. {\dzero} has an
upper cut on {\met} of 200~GeV to reject misreconstructed events. Both
collaborations throw out events with low {\met} just above the cut
thresholds when it is aligned or back-to-back with one of the objects
in the event, indicative of a misreconstructed event. The primary
vertex must be clearly identified and near the center of the detector,
and the lepton must originate from it. The regions between {\dzero}'s
central and end calorimeter cryostats are tricky to instrument and
model accurately, and if the leading jet in the muon analysis channel
points to this region, the threshold on it is raised to
30~GeV. Finally, {\dzero} has cuts on muon track curvature significance
designed to reject events where the muon has been misreconstructed. In
CDF's {\met}+jets analysis, a neural network with 15 input variables
is trained to separate the multijets background from signal and a cut
is placed on the output distribution.

After the kinematic event selection, {\dzero}'s background samples
retain 4~million MC events and 0.8~million multijet data events, and
there are 0.5 million single top quark signal MC events. The signal
data contain 114,777 events, with predicted background components:
$Wjj = 71\%$, $Wcj = 6\%$, $W{\ccbar} = 6\%$, $W{\bbbar} = 3\%$,
$Z$+jets = 6\%, dibosons = 2\%, {\ttbar} = 1\%, and multijets =
5\%. The expected single top quark signal is $tb = 0.13\%$ and $tqb =
0.26\%$, with a signal-to-background ratio for $tb$+$tqb$ of
1:260. Clearly, an additional method is needed to select events for
the analyses to stand any chance of finding the single top quark
signal.

\subsection{Heavy-Flavor Jet Tagging}

The most powerful part of event selection is the identification of
jets that originate from $b$~quarks. The algorithms use the long decay
time of the $B$~hadrons (mean lifetime $\simeq 1.5 \times 10^{-12}$~s)
which results in detached secondary vertices in the jets ($> 1$~mm
between the primary and secondary vertices), together with other
information about the tracks to find $b$~jets. The tagging algorithms
are applied directly to jets in data and to most MC events at CDF, and
are modeled with tag-rate functions for MC events at {\dzero} together
with taggability-rate functions to reproduce the detector geometric
acceptance and operating efficiency. For $W$+light jets MC, CDF uses
tag-rate functions measured in multijets data. $b$-jet identification
is implemented at {\dzero} by combining all the track and vertex
information using a neural network.\cite{btaggingdzero} CDF uses the
significance of the decay length of the secondary vertex in the
($r,\phi$) plane for the lepton+jets and {\met}+jets
channels,\cite{btaggingcdf1} and also a jet probability algorithm in
the {\met}+jets channel.\cite{btaggingcdf2} Depending on where a cut
is put on these variables, one can define looser or tighter
$b$~tagging, where ``loose'' means higher probability to tag a $b$~jet
(58\% for $b$~jets within {\dzero}'s Silicon Microstrip Tracker
fiducial geometric acceptance) with associated higher probability to
mistag a non-$b$ jet (17\% for charm jets and 1.8\% for light quark
and gluon jets), and ``tight'' means a lower $b$-tag probability (47\%
for $b$~jets at {\dzero}) with associated lower fake tag rates (10\%
for $c$~jets and 0.5\% for light jets). {\dzero} requires one
tight-tagged jet (and no loose-tagged jet) for its single-tagged
analysis channels, and two loose-tagged jets for its double-tagged
channels. CDF has one set point for both single-tagged and
double-tagged lepton+jets channels, with efficiencies of 50\% ($b$),
9\% ($c$), and 1\% ($j$) for fiducial jets within the Silicon Detector
tracking system.

\subsection{Analysis Channel Separation}

To improve the sensitivity of the measurement, both {\dzero} and CDF
split their datasets into independent channels using the jet
multiplicity (2, 3; and 4 for {\dzero}), number of $b$-tagged jets (1
or 2), lepton flavor ({\dzero} only, electron or muon), trigger type
(CDF only, lepton, {\met} for muon+jets) and data-collecting period
({\dzero} only, Run~IIa and Run~IIb), giving 24 independent
lepton+jets analysis channels for {\dzero} and eight for CDF. CDF's
{\met}+jets channel with no isolated lepton is split by the number and
type of $b$~tags (one SecVtx-tagged jet, two ``SecVtx''-tagged jets,
and one ``SecVtx'' and one ``JetProb''-tagged jet). Measurements are
made in each channel and combined at the end of the analysis. The
signal-to-background ratios vary from 1:10 (2-jets/2-tags) to 1:37
(4-jets/2-tags) for {\dzero}, with the most important 2-jets/1-tag
channels having S:B = 1:20. CDF's channels have S:B = 1:15 in the
2-jets channels (1-tag and 2-tag combined), S:B = 1:23 in the 3-jets
channels, and S:B = 1:23 in the {\met}+jets channels.


\section{Signal Acceptances and Event Yields}

After all event selections have been applied, the signal acceptances
(percentage of total cross section) for {\dzero} are $(3.7 \pm 0.5)\%$
for the s-channel $tb$ process and $(2.5 \pm 0.3)\%$ for the t-channel
$tqb$ process. The t-channel process has a lower acceptance because
the $\bar{b}$ jet has low transverse momentum and is difficult to
identify. CDF's signal acceptances in the lepton+jets channels are
2.7\% for the $tb$ process and 1.8\% for the $tqb$ process. These
values are lower than {\dzero}'s because of the more restrictive
trigger requirements, tighter kinematic selection, and tighter
$b$~tagging in the double-tagged channel. In addition, CDF has the
{\met}+jets channel with a signal acceptance of 1.1\% for $tb$+$tqb$
combined.

Table~\ref{event-yields} shows the numbers of signal and background
events expected, and the numbers of data events found. For simplicity
here, all analysis channels have been combined. Four notes to
understand the table are in order: (i) Remember that {\dzero} uses
$m_t = 170$~GeV and CDF uses 175~GeV for single top signal and
{\ttbar} background, with associated higher theory cross sections for
the lower top quark mass. They each also use different theory
calculations for these values: for single top, {\dzero} uses Kidonakis
2006 values of $1.12
\pm 0.05$~pb ($tb$) and $2.34 \pm 0.13$~pb
($tqb$),\cite{singletopkidonakis} and for {\ttbar} they use the
Kidonakis and Vogt 2003 value of $7.91^{+0.61}_{-1.01}$~pb (where the
{\ttbar} uncertainty includes a component for the top quark
mass).\cite{ttbarxsec2003} CDF uses for single top the Harris {\it et
al.} 2002 values of $0.88 \pm 0.12$~pb ($tb$) and
$1.98^{+0.28}_{-0.22}$~pb ($tqb$),\cite{singletopharris} and for
{\ttbar} they use the Bonciani {\it et al.} 1998 value of $6.70 \pm
1.32$~pb.\cite{ttbarxsec1998} Thus, direct comparison of the signals
and {\ttbar} backgrounds needs one or other experiment's numbers to be
rescaled to be valid. (ii) {\dzero}'s analysis includes channels with
four jets and CDF's does not, so the fraction of {\ttbar} events
expected by {\dzero} is higher than at CDF when showing yields with
all channels combined. However, when one considers each jet
multiplicity channel separately, then the relative fractions of
$W$+jets, {\ttbar}, etc. are very similar between the two
experiments. (iii) CDF's {\met}+jets channel $W$+jets yield does not
include $Wjj$ where $j$ = a light jet. (iv) CDF's {\met}+jets channel
multijets yield includes also the $Wjj$ events.

\begin{table}[!h!tbp]
\tbl{Numbers of events after all selections have been applied. See
comments in the text on how to compare the columns.
\label{event-yields}}
{\begin{tabular}{@{}lccc@{}}
\toprule
       & {\dzero}'s Yields    & \multicolumn{2}{c}{CDF's Yields} \\
       & Lepton+Jets, 2.3~fb$^{-1}$ & Lepton+Jets, 3.2~fb$^{-1}$
                                    & {\met}+Jets, 2.1~fb$^{-1}$ \\
\colrule
$tb$+$tqb$ signal  &   $223 \pm 30$  &   $191 \pm 28$  &    $64 \pm 10$  \\
$W$+jets           & $2,647 \pm 241$ & $2,204 \pm 542$ &   $304 \pm 116$ \\
$Z$+jets, dibosons &   $340 \pm 61$  &   $171 \pm 15$  &   $171 \pm 54$  \\
{\ttbar} pairs     & $1,142 \pm 168$ &   $686 \pm 99$  &   $185 \pm 30$  \\
Multijets          &   $300 \pm 52$  &   $125 \pm 50$  &   $679 \pm 28$  \\
\colrule
Total prediction   & $4,652 \pm 352$ & $3,377 \pm 505$ & $1,403 \pm 205$ \\
Data               &       4,519     &       3,315     &       1,411     \\
\botrule
\end{tabular}}
\end{table}

\begin{figure}[!h!tbp]
\begin{center}
\epsfig{file=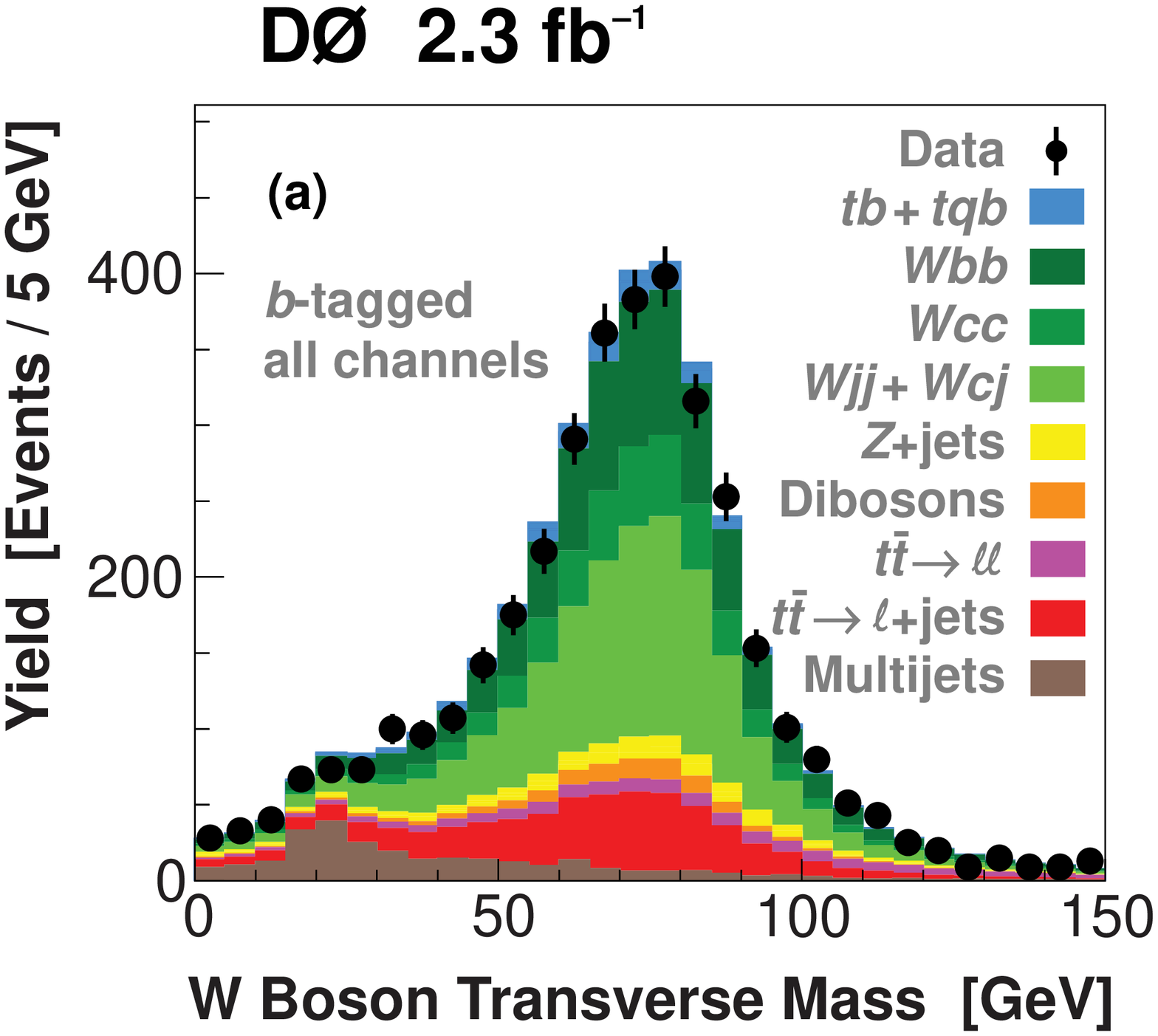,width=2.38in}
\hspace{0.1in}
\epsfig{file=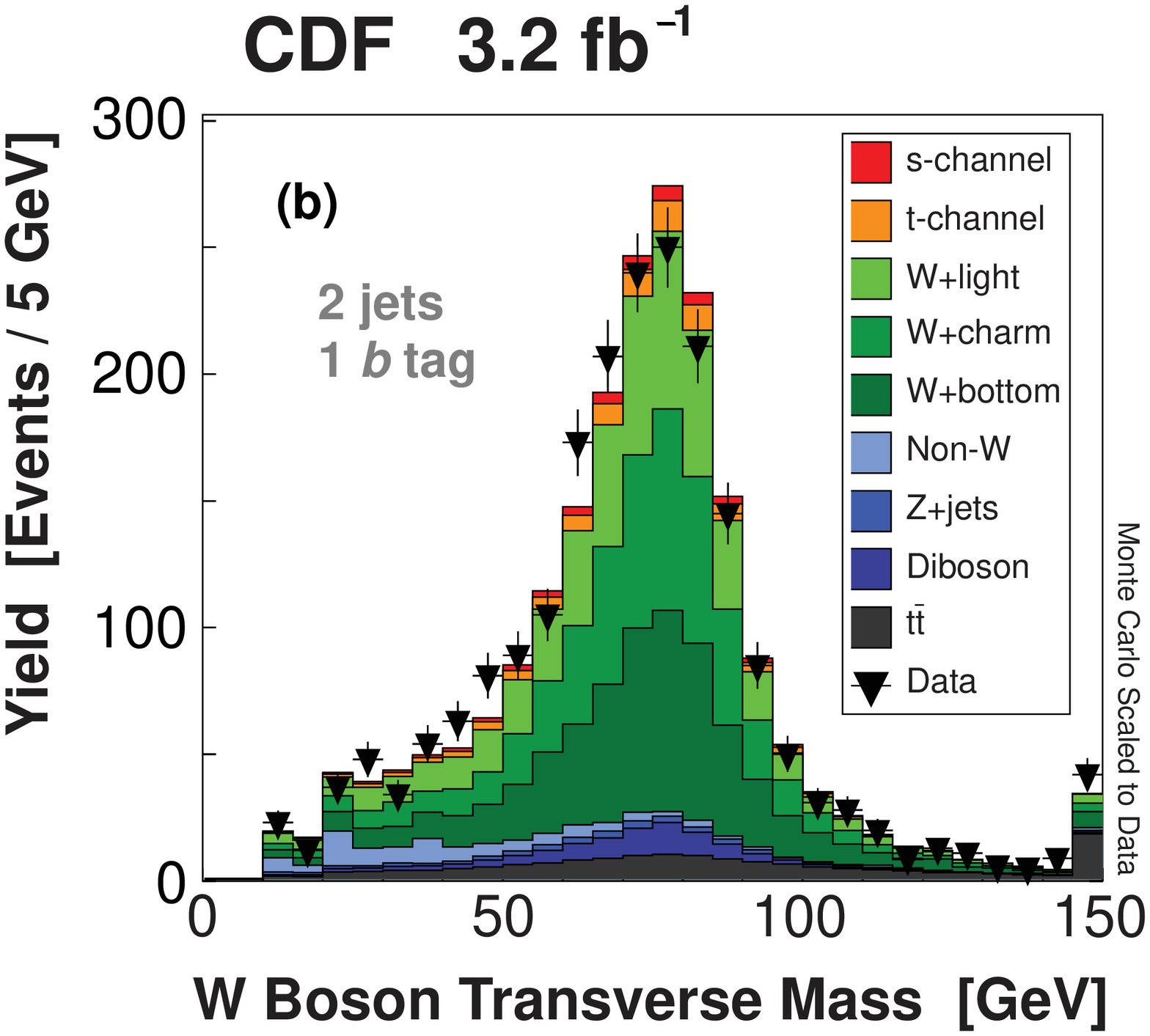,width=2.38in}
\end{center}
\vspace{-0.1in}
\caption[MTW]{Distributions of the $W$~boson transverse mass for (a)
{\dzero}, with all analysis channels combined, and (b) CDF, with all
lepton+2-jets/1-tag channels combined. \label{W-transverse-mass}}
\end{figure}

Figure~\ref{W-transverse-mass} shows the reconstructed $W$~boson
transverse mass distributions from {\dzero} (all channels combined)
and CDF (lepton+2jets channels). The transverse mass is defined as:
$M_T(W) = M_T(l,\nu) =
\sqrt{2p_T(l){\met}(1-\cos(\phi(l)-\phi({\met})))}.$

\section{Background Model Checks}

\begin{figure}[!h!tbp]
\begin{center}
\epsfig{file=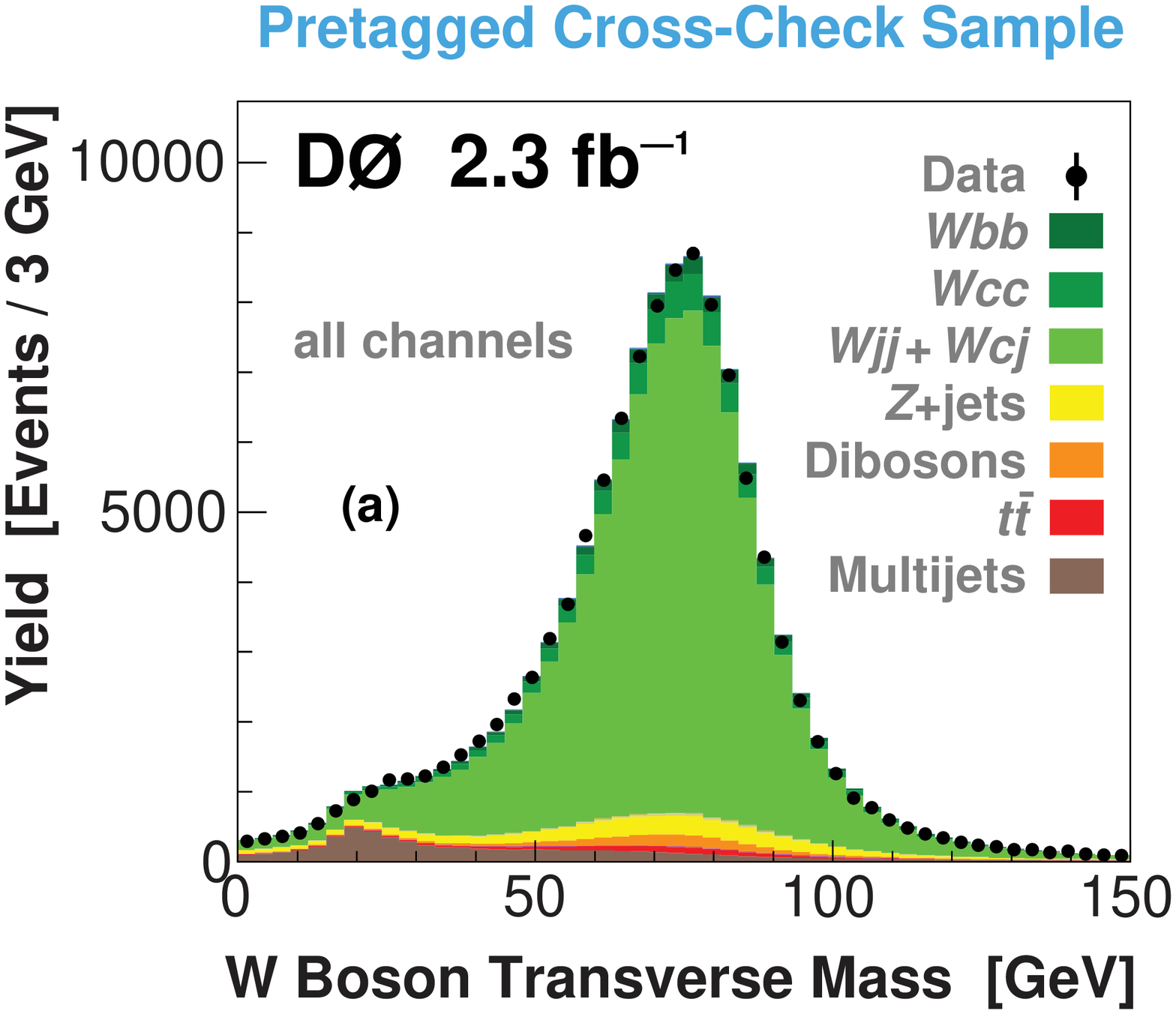,width=1.56in}
\hspace{0.075in}
\epsfig{file=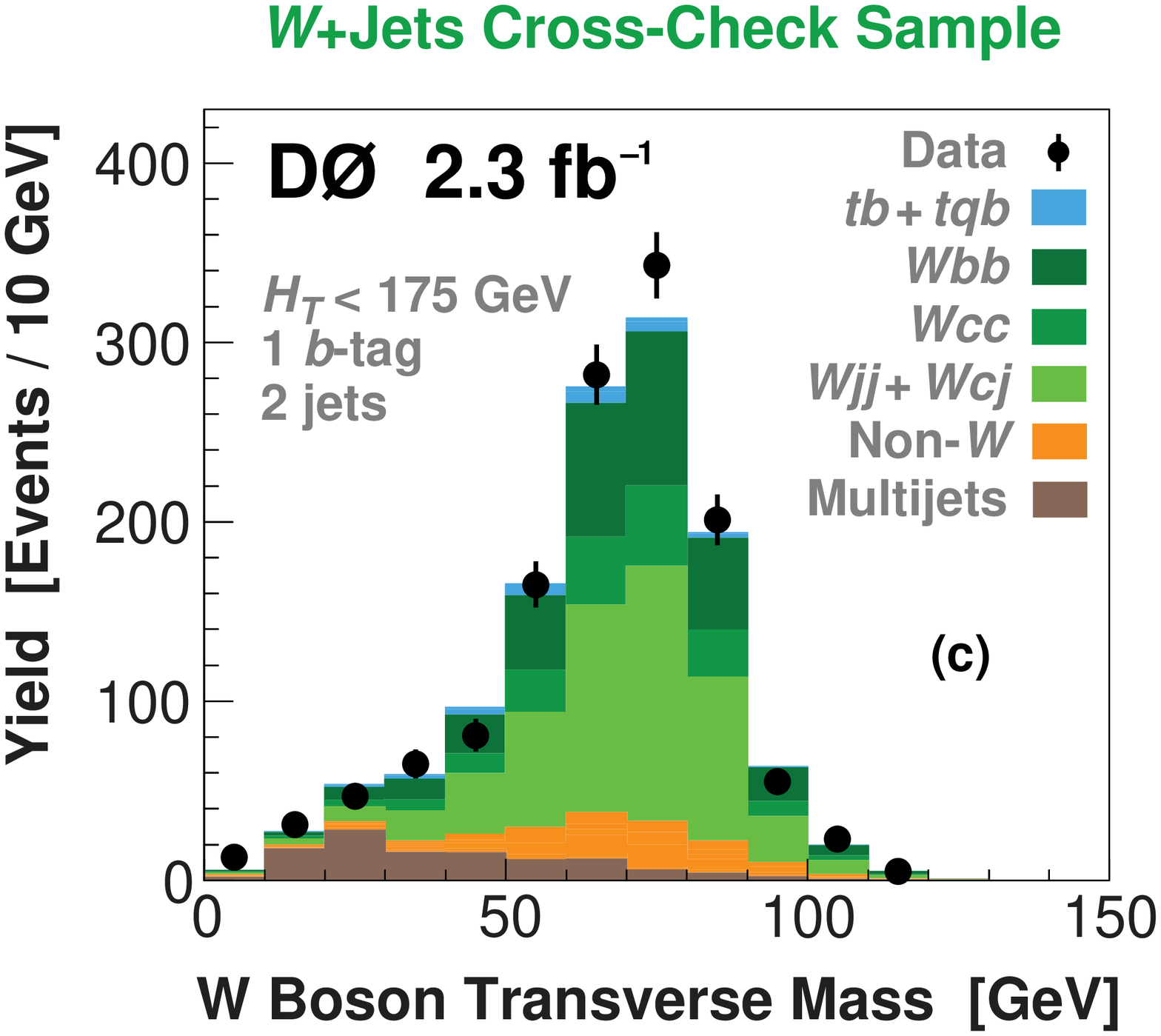,width=1.56in}
\hspace{0.075in}
\epsfig{file=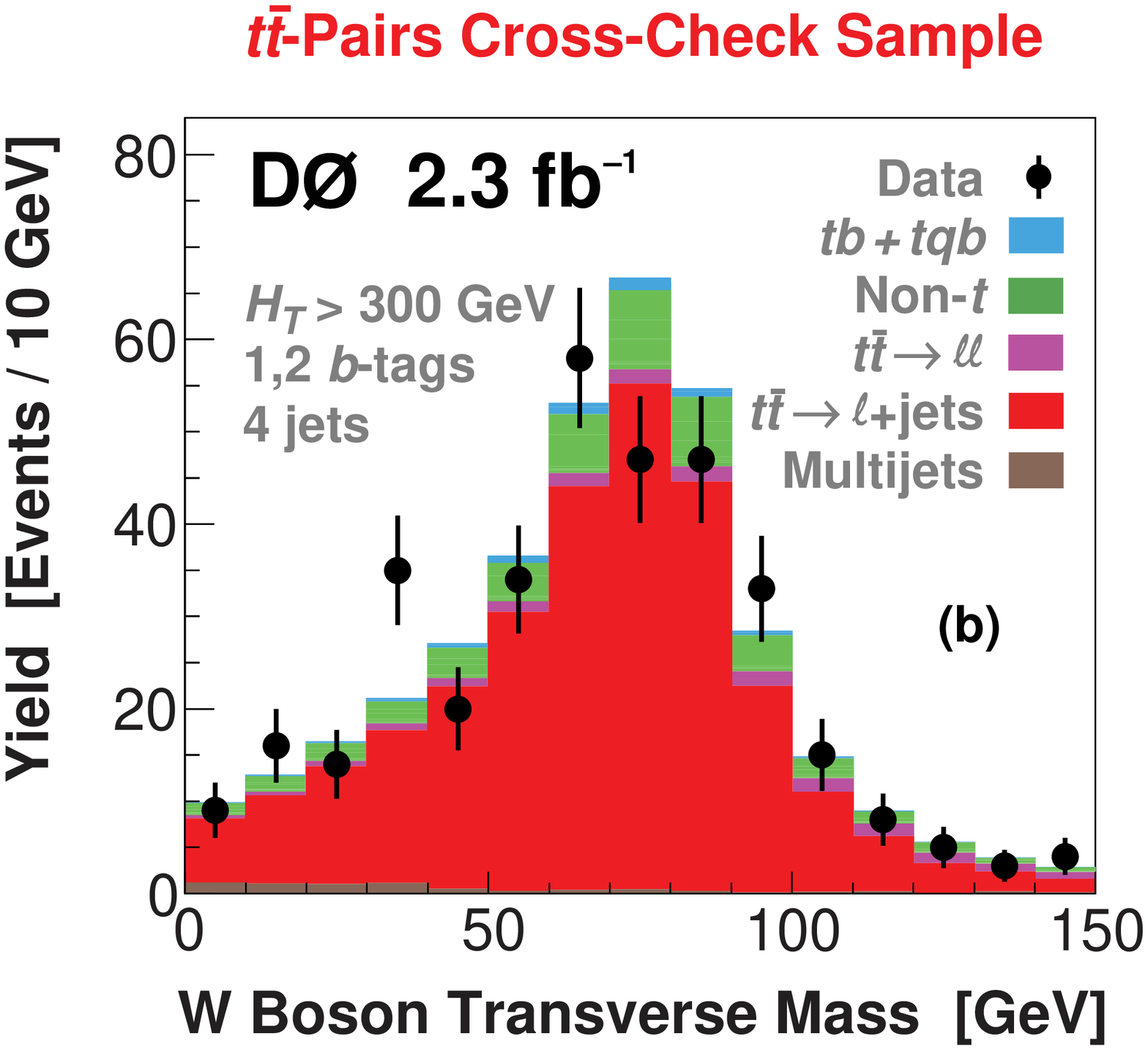,width=1.56in}
\\
\epsfig{file=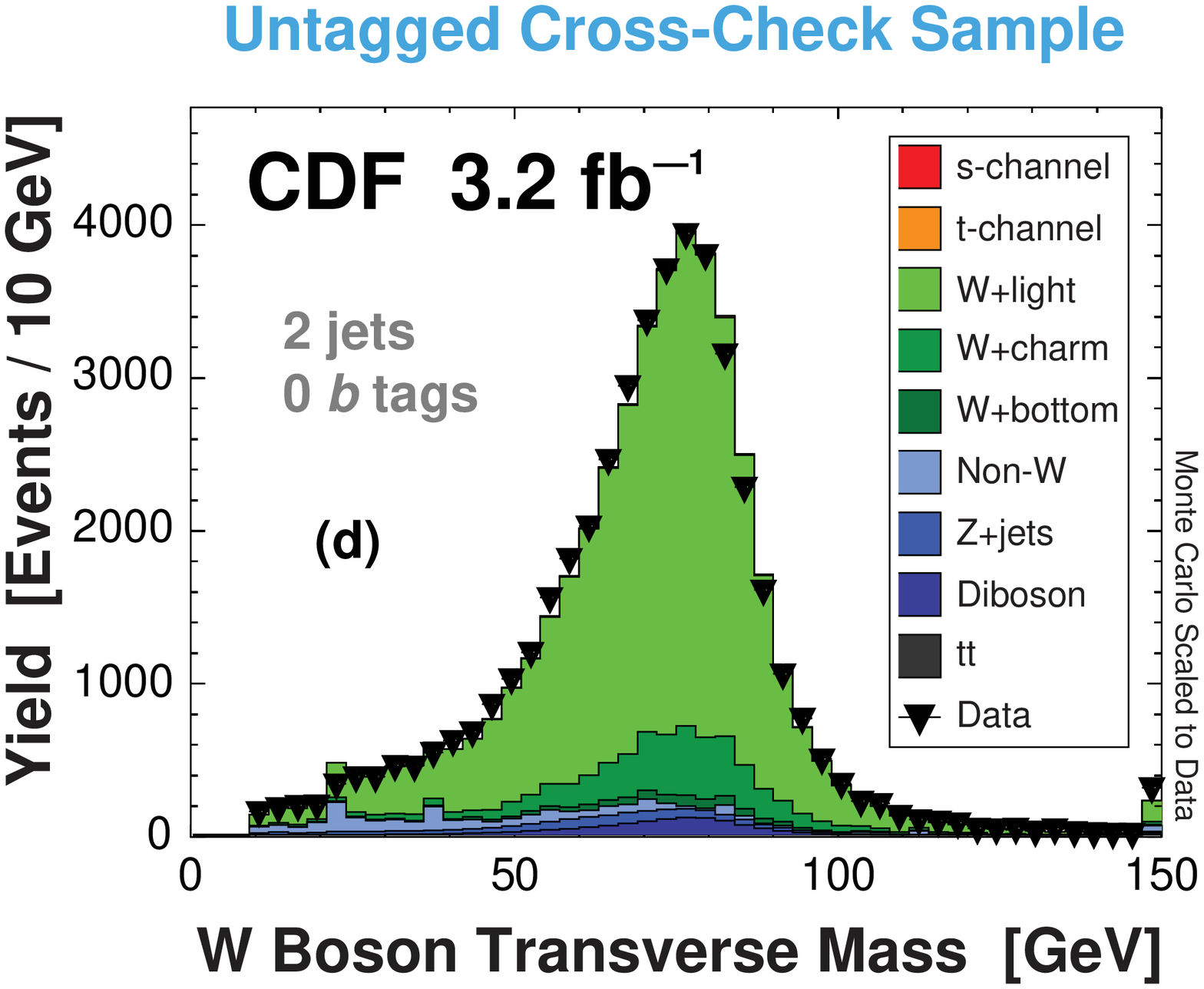,width=1.65in}
\hspace{0.075in}
\epsfig{file=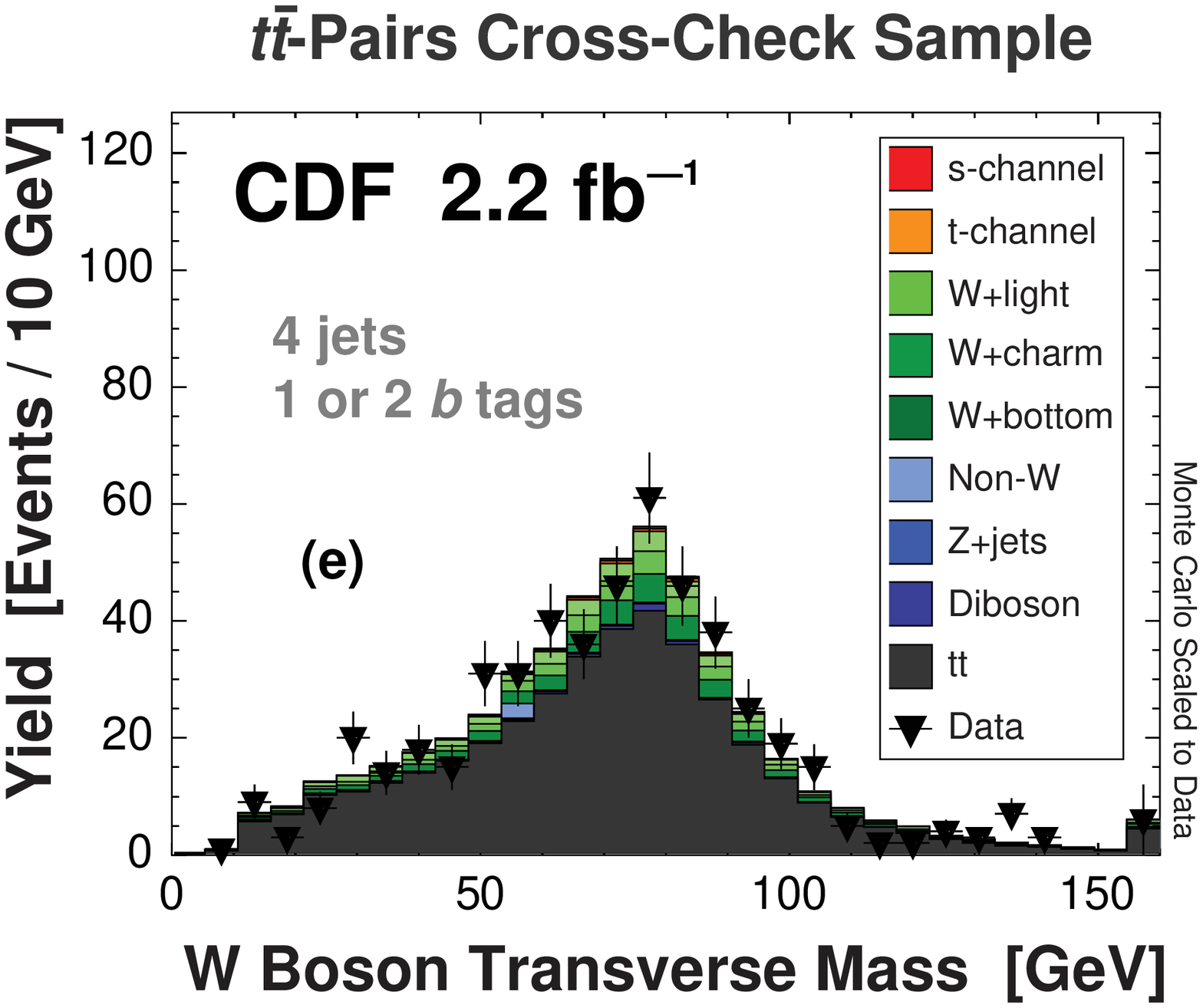,width=1.65in}
\end{center}
\vspace{-0.1in}
\caption[crosscheck]{Distributions of the $W$~boson transverse mass
for several cross-check samples: (a)~{\dzero}'s pretagged events, with
all analysis channels combined, (b)~{\dzero}'s $W$+jets cross-check
sample, (c)~{\dzero}'s {\ttbar} pairs cross-check sample, (d)~CDF's
two-jets untagged sample, and (e)~CDF's {\ttbar} cross-check
sample. \label{cross-checks}}
\end{figure}

Any analysis of the data is only valid if the background models
reproduce the data in all variables used for event selection and to
separate signal from background. In addition to checking the
background model agreement with data for these distributions for every
particle in each analysis channel, extensive cross checks using other
data samples have been performed to ensure the separate components of
the background model are accurately modeled. Samples that pass all
selection cuts are used before $b$~tagging to certify the shape of the
$W$+light jets background model. The $W$+heavy flavor background
model's agreement between data and background model in both shape and
normalization is checked using a sample with exactly two jets, with
one $b$~tagged, and $H_T(\ell,{\met},{\rm jets}) < 175$~GeV in
{\dzero}'s analysis. Finally, the {\ttbar} background is validated in
both normalization and shape using data and MC samples with four jets,
one or two $b$~tags, and, for {\dzero} only since they have softer
object $E_T$ requirements, $H_T(\ell,{\met},{\rm jets}) >
300$~GeV. Many distributions are checked using these three cross-check
samples and good agreement between data and background model is
found. Figure~\ref{cross-checks} shows the transverse mass of the
$W$~boson as an example.

\section{Systematic Uncertainties}

The uncertainties in all searches are dominated by the statistical
uncertainty from the size of the data sample. However, once there is
enough data to observe and measure a signal, then systematic
contributions to the total uncertainty become important. The total
uncertainty on the cross section measurement by {\dzero} is $\pm22\%$,
and for CDF it is $+29\%, -24\%$ in the lepton+jets channel, $+52\%,
-46\%$ in the {\met}+jets channel, and $+26\%, -22\%$ with these
channels combined. The contribution from the data statistics in
{\dzero}'s measurement is $\pm18\%$, leaving $\pm13\%$ from systematic
components. Normalization systematic uncertainties and shape-dependent
systematic uncertainties are considered separately for each signal and
background source in each analysis channel. The overall background
uncertainty varies between 7\% and 15\% for the individual channels in
{\dzero}'s measurement. Shape and normalization uncertainties combined
result in 20\% uncertainties on the background model for single-tagged
channels and 40\% uncertainties on background for double-tagged
channels, for events most like signal. The uncertainties on the
background model for events most like background about 10\% for
single-tagged channels and 15\% for double-tagged channels. {\dzero}
measures systematic uncertainty contributions from 23 different
sources. Others were considered but found to be negligible. The
largest source of systematic uncertainty comes from the $b$-ID
tag-rate functions, including both normalization and shape parts,
followed by the jet energy scale calibration (also normalization and
shape), and the heavy-flavor correction factor for the $W{\bbbar}$ and
$W{\ccbar}$ fractions in the MC model. Smaller contributions (in
descending order) come from the integrated luminosity, the jet energy
resolution, initial-state and final-state radiation, $b$-jet
fragmentation, the {\ttbar} pairs cross section, and lepton
identification. CDF's analyses include normalization uncertainty terms
from 16 sources, and shape terms from a subset of nine sources. The
most important ones are the jet energy scale, the event detection
efficiency, and the $W{\bbbar}$, $W{\ccbar}$, and $Wcj$ scale factor.


\section{Signal-Background Separation}

The sensitivity to observe a signal with a large background is greatly
improved by finding a variable that has a different shape for signal
than for background. One can then keep only events in the
maximal-signal region and measure a cross section by counting events
if there is enough data, or, as in the case of single top quark
production with only a few inverse femtobarns of data, one can perform
a binned likelihood calculation comparing the shapes of the expected
signal and background to data across the full distribution to further
improve the sensitivity. Since the kinematics of single top quark
events lie between those of the dominant lower-energy $W$+jets and
higher-energy {\ttbar} backgrounds, it is not possible to find a
single simple variable with which to perform this calculation. Hence,
{\dzero} and CDF each combine many variables using several different
methods to increase the signal-background separation power. {\dzero}
uses three discrimination methods and CDF uses five in the lepton+jets
channel, one in a separate s-channel $tb$ search, and one in the
{\met}+jets channel, which are briefly described here; more details
are available
elsewhere.\cite{singletopdzero6}$^,$\cite{webpagedzero}$^,$\cite{webpagecdf}

\subsection{Discriminating Variables}

{\dzero} uses 97 discriminating variables in its final analysis,
chosen from a much longer list to include only those variables with a
different distribution for signal and at least one of the background
components,\cite{singletopyuan}$^,$\cite{singletopmahlon}$^,$\cite{singletopcau1}$^,$\cite{singletopcau2}$^,$\cite{liu}
and also to have good agreement between the shape of the background
sum and data. The variables fall in five categories: object
kinematics, event kinematics, jet reconstruction, top quark
reconstruction, and angular correlations. The most powerful ones for
separating single top quark signal from the $W$+jets and {\ttbar}
backgrounds in each category are shown in Table~\ref{variables}.

\begin{table}[!h!tbp]
\tbl{30 of the 97 variables used by {\dzero} that have the best
separation between the single top quark signal and $W$+jets or {\ttbar}
pairs. \label{variables}}
{\begin{tabular}{@{}lcc@{}}
\toprule
               & \multicolumn{2}{c}{Separate Single Top from:}      \\
Variable Type  & $W$+Jets                        & {\ttbar} Pairs  \\
\colrule
Object         & {\met}                          & $p_T$(notbest2) \\
Kinematics     & $p_T$(jet2)                     & $p_T$(jet4)     \\
               & $p_T^{\rm rel}$(jet1,tag-$\mu$) & $p_T$(light2)   \\
               & $E$(light1)                     &                 
\vspace{0.1in} \\
Event          & $M$(jet1,jet2)                  & $M({\rm alljets}-{\rm tag1})$   \\
Kinematics     & $M_T(W)$                        & Centrality(alljets)             \\
               & $H_T$(lepton,{\met},jet1,jet2)  & $M({\rm alljets}-{\rm best1})$  \\
               & $H_T$(jet1,jet2)                & $H_T({\rm alljets}-{\rm tag1})$ \\
               & $H_T$(lepton,{\met})            & $H_T({\rm lepton},{\met},{\rm alljets})$ \\
               &                                 & $M$(alljets)                    
\vspace{0.1in} \\
Jet            & Width$_{\phi}$(jet2)            & Width$_{\eta}$(jet4) \\
Reconstruction & Width$_{\eta}$(jet2)            & Width$_{\phi}$(jet4) \\
               &                                 & Width$_{\phi}$(jet2) 
\vspace{0.1in} \\
Top Quark      & $M_{\rm top}(W({\rm S1}),{\rm tag1})$     &                       \\
Reconstruction & $\Delta M_{\rm top}^{\rm min}$            &                       \\
               & $M_{\rm top}(W({\rm S2}),{\rm tag1})$     &                       
\vspace{0.1in} \\
Angular        & $\cos$(light1,lepton)$_{\rm btaggedtop}$  & $\cos$(lepton$_{\rm btaggedtop}$,btaggedtop$_{\rm CM}$) \\
Correlations   & $\Delta\phi$(lepton,{\met})               & $Q({\rm lepton})\times\eta({\rm light1})$ \\
               & $Q({\rm lepton})\times\eta({\rm light1})$ & $\Delta R$(jet1,jet2) \\
\botrule
\end{tabular}}
\end{table}

Some comments on the notation are in order. The numbering $n$ of
jet$n$, tag$n$, light$n$, etc. refers to the transverse momentum
ordering of the jets, 1 is the highest $p_T$ jet of that type of jet,
2 is the second-highest $p_T$ jet, and so on. ``tag'' means a
$b$-tagged jet. ``light'' means an untagged jet (it fails the $b$-tag
criteria). ``best'' means the jet which, when combined with the lepton
and missing transverse energy, produces a top quark mass closest to
170~GeV (the value at which {\dzero}'s analysis is
performed). ``notbest'' means any jet that is not the best
jet. ``alljets'' means include all the jets in the event in the global
variable. $p_T$ is the transverse momentum. $E$ is the particle's
energy. $Q$ is the particle's charge. $H_T$ is the scalar sum of the
particles' transverse energies. $M$ is the invariant mass of the
objects. $M_T$ is the transverse mass of the objects. $p_T^{\rm rel}$
is the transverse momentum of the muon closest to the jet relative to
that jet. S1 and S2 are the two solutions for the neutrino
longitudinal momentum when solving the $W$~boson mass equation, and S1
is the smallest absolute value of the two (the preferred
value). $\Delta M_{\rm top}^{\rm min}$ is the difference between
170~GeV and the reconstructed top quark mass using the jet and
neutrino solution that make the mass closest to 170~GeV. $\Delta
R({\rm object1}, {\rm object2}) = \sqrt{\Delta
\phi({\rm object1}, {\rm object2})^2 + \Delta \eta({\rm object1}, {\rm
object2})^2}$. Finally, subscripted text in the cosines indicates the
rest frame in which to measure the variable in question. ``CM'' is the
center of mass frame of the whole final state.

The CDF collaboration uses fewer variables with their discriminants,
but they have one very powerful variable not developed by {\dzero}:
the jet flavor separator.\cite{lueck} This takes all parameters that
describe the tracks in $b$-tagged jets and combines them using a
neural network to calculate a probability that the jet is a bottom
jet, or a charm or light quark or gluon jet. This variable increases
the signal-background separation sensitivity by 15\%.

\subsection{Boosted Decision Trees}

A decision tree\cite{decisiontrees} applies sequential cuts to the
events but does not reject events that fail the cuts. The choice of
variables and cuts at each level of the tree is made by training the
trees on large sets of signal and background MC events.
Boosting\cite{boosting} averages the results over many trees and
improves the performance by about 20\%. {\dzero} pioneered the use of
boosted decision trees (BDTs) to separate signal from background in
the single top search in
2006.\cite{singletopdzero5}$^,$\cite{singletopdzero6} They use custom
code with 64 input variables from the total list of 97, and 50
boosting cycles with a separate set of BDTs for each of the 24
analysis channels.\cite{gillberg}$^,$\cite{benitez} The same variables
are used in every analysis channel, since the BDTs ignore ones that do
not show sensitivity in any particular channel. With BDTs, there is
also no need to split the signal and background samples by
subcomponent to improve the sensitivity (which is beneficial with
traditional neural networks\cite{singletopdzero2}), since they handle
the varying kinematics without problem. The CDF collaboration also
uses BDTs, recently included in the {\sc tmva} package in {\sc
root}.\cite{tmvaroot} They use 22 variables for 2-jet events and 29
for 3-jet events, with 400--600 boosting cycles.\cite{casal} They
train four sets of BDTs in total, since they combine electron and muon
channels and the two trigger types. After boosting, the distributions
of both signal and background are highly centralized between zero and
one. In order to avoid using bins in the final calculation with
predicted signal or data but no predicted background, {\dzero}
transforms its output distributions (from all three discriminant
methods, not just BDTs) to ensure that every bin has at least 40
background events. This transformation clusters the background events
near zero and the signal events near one, and avoids instabilities in
the final cross section measurement.

\subsection{Traditional Neural Networks}

{\dzero} made the first particle search using neural networks (NNs) to
separate signal from background in 2001.\cite{singletopdzero2} The
type used were multilayer feed-forward perceptrons from the {\sc
mlpfit} package.\cite{mlpfit} CDF uses NNs in the observation analysis
for the lepton+jets channels with 14 input variables,\cite{lueck} and
for the {\met}+jets channels with 11 variables.\cite{tmvaroot} The
networks in the lepton+jets channels come from the commercial {\sc
neurobayes} package.\cite{neurobayes} Despite its name, it is not a
Bayesian NN package as described in the next subsection. The networks
are trained on the same events as used with the BDTs to obtain the
weights between nodes and thresholds at the nodes. An independent set
of events is used to test the networks after each training cycle to
avoid overtraining. Since NNs use all input variables (unlike BDTs,
which ignore ones not found to be useful), care must be taken not to
include variables with insufficient separation power uncorrelated from
the other variables, otherwise noise is introduced into the system and
the separation can decrease. This is the reason why far fewer
variables are used with NNs than with BDTs.

\subsection{Bayesian Neural Networks}

{\dzero} introduced the use of Bayesian neural networks
(BNNs)\cite{bayesianNNs} for signal-background separation in the 2006
single top evidence
analysis.\cite{singletopdzero5}$^,$\cite{singletopdzero6} Like
traditional NNs, a short list of input variables must be chosen, and
{\dzero} uses the {\sc rulefit} package\cite{rulefit} to select
between 18 and 28 variables per analysis channel. The networks have 20
hidden nodes. The Bayesian part of this technique is to average over
many networks in each channel using the Markov-Chain MC sampling
technique.\cite{mcmc} {\dzero} uses 300 networks in each of the 24
analysis channels, with the final result in each channel being taken
from an average of the last 100 networks in the
chain.\cite{tanasijczuk} This averaging process makes the
discrimination insensitive to details of which events are used in
training, so it is not possible to overtrain the networks, although
closure tests are performed using independent events to verify
convergence. It is also not necessary to split the signal and
background components with separate networks for optimal
separation. The averaging process also improves the signal-background
separation, since it is not dependent on the choice of starting
parameters for the weights between nodes or thresholds at the nodes,
which can lead to solutions at local minima which are not optimal
without the averaging.

\subsection{Matrix Elements}

Both {\dzero} and CDF use matrix elements (MEs) to separate signal
from background.\cite{pangilinan}$^,$\cite{dong} {\dzero} developed
the method to measure the top quark mass in 2004,\cite{naturetopmass}
and was the first to apply them to signal-background separation in the
2006 single top evidence
analysis.\cite{singletopdzero5}$^,$\cite{singletopdzero6} The matrix
elements correspond to signal and background probability
densities. {\dzero} calculates matrix elements for three signal
processes in the 2-jets channel and five in the 3-jets channel,
together with eight background processes in the 2-jets channel and
three in the 3-jets channel. The proton and antiproton are modeled
using parton distribution functions and detector resolutions are taken
into account using jet resolution transfer functions. The calculations
are extremely CPU-intensive, and take many months to complete. To
improve the sensitivity, {\dzero} splits the analysis into events with
$H_T < 175$~GeV (mainly $W$+jets background) or $H_T \ge 175$~GeV
(mainly {\ttbar} and hard $W$+jets background).

\subsection{Likelihood Functions}

CDF has an analysis that uses likelihood functions for
signal-background separation with $tb$+$tqb$ as signal.\cite{budd}
They also search separately for only the s-channel $tb$ process, using
different likelihood functions and input variables.\cite{jung}
Likelihoods are much simpler than NNs, they need no training on signal
or background event samples, and do not take correlations between the
variables into account. In the 2-jet channels, CDF's likelihoods
combine seven variables, including two powerful ones: the logarithm of
the matrix element, and the jet flavor separator. In the 3-jet
channels, 10 variables are combined.

\subsection{Combining the Discriminant Outputs}

\begin{figure}[!h!tbp]
\begin{center}
\epsfig{file=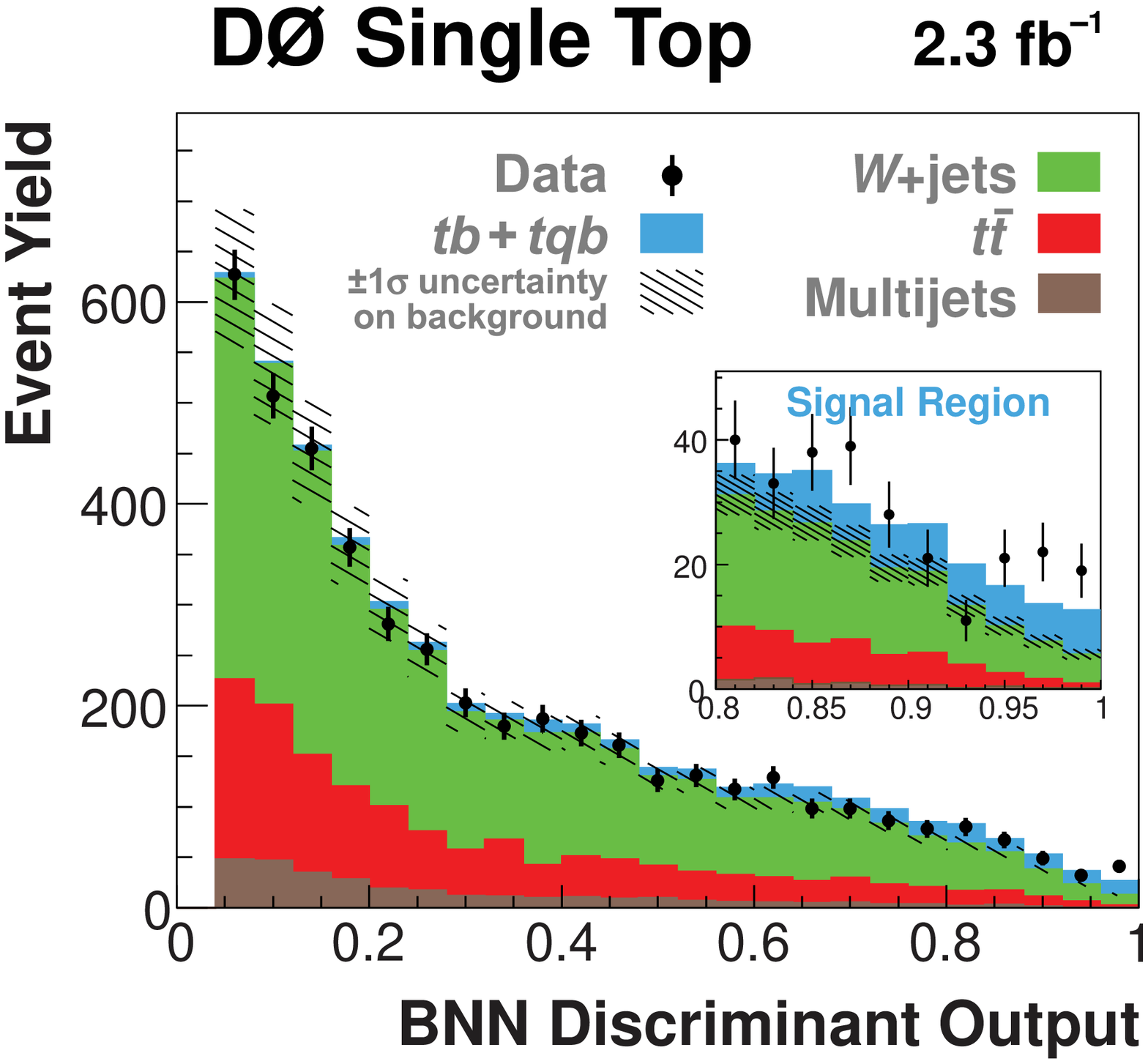,width=2.38in}
\hspace{0.1in}
\epsfig{file=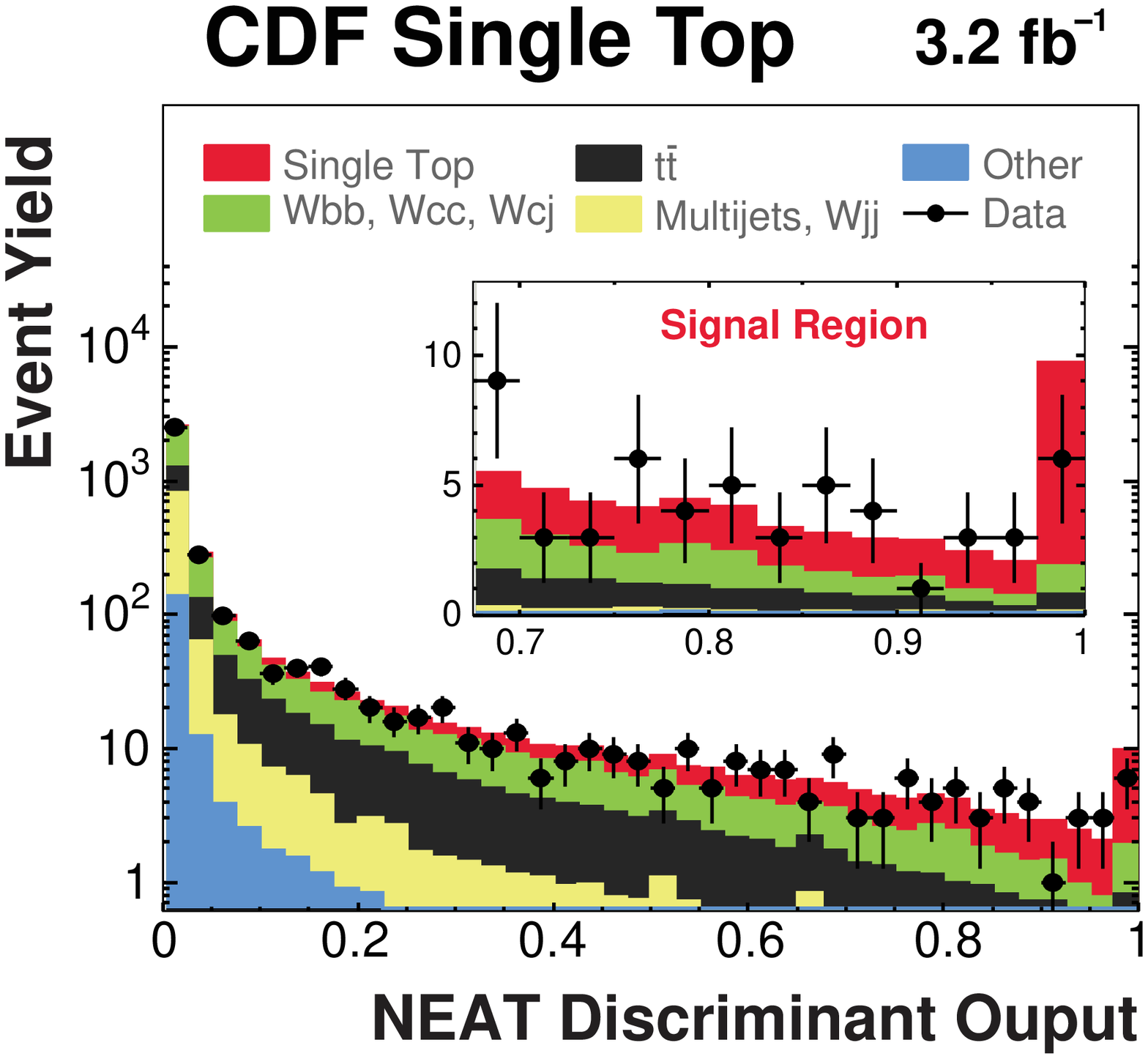,width=2.38in}
\end{center}
\vspace{-0.1in}
\caption[final-outputs]{Output distributions from (a)~{\dzero}'s BNN
combination discriminant, for all analysis channels combined, and
(b)~CDF's NEAT combination discriminant, for all lepton+jets analysis
channels combined. \label{final-outputs}}
\end{figure}

The measurements from each discrimination method are correlated, but
by less than 100\%, and the discriminant outputs may thus be combined
to improve the precision of the final measurement. {\dzero} measures
the correlation between its three analysis methods (BDT, BNN, ME)
using an ensemble of pseudodatasets containing background and SM
signal, and finds the correlation to be 74\% between BDT and BNN, 60\%
between BDT and ME, and 57\% between BNN and ME. To combine the three
measurements in each of the 24 analysis channels, {\dzero} uses an
additional set of BNNs, each with three inputs and six hidden
nodes. CDF uses an innovative method to combine its lepton+jets
measurements: neuro-evolution of augmenting technologies
(NEAT).\cite{NEAT} This is a method for evolving neural networks with
a genetic algorithm. Evolution starts with small simple networks that
become increasingly complex over sequential generations. The networks
are trained to give the best expected p-value (significance) for the
result. This is unlike how traditional NNs are optimized during
training, when the error function (signal-background similarity) is
minimized. The NEAT networks are also used to optimize the binning for
the measurement. Figure~\ref{final-outputs} shows the final output
distributions for all analysis channels combined.

\vfill
\clearpage

\vspace{-0.2in}

\section{Cross Section Measurements}

\subsection{Bayesian Binned Likelihoods}

The distributions from the combination discriminants from 24
independent lepton+jets analysis channels at {\dzero} and eight
lepton+jets plus three {\met}+jets channels at CDF are used in a
Bayesian binned likelihood calculation to extract the single top quark
cross section. A flat nonnegative prior is used for the signal cross
section. All systematic uncertainties on background normalization and
shape and signal acceptance and their correlations are taken into
account. The shape uncertainties from the jet energy scale are
smoothed from bin to neighboring bin during the calculation. Using the
full range of the discriminant outputs for this calculation means that
the high statistics background-dominated region (near zero) is used to
constrain the uncertainties on the much smaller background in the
expected-signal-dominated region (near one). The signal cross section
central value is taken from the position of the peak of the posterior
density distribution, and the uncertainty on the cross section
(statistical and systematic components combined) comes from the width
of the distribution about the peak that encompasses 68\% of its area
($\pm 1 \sigma$). The cross section calculations are also performed
using the outputs from each discriminant method separately, and using
subsets of the data (all electron+jets channels, all 2-jets channels,
all 1-tag channels, and so on) to check for consistency, which is
found within the statistical uncertainty on the measurements.

\vspace{-0.1in}

\subsection{Ensembles and Linearity Studies}

To check that the discriminants do not introduce a bias into the
measured cross section, {\dzero} generates eight ensembles of
pseudodatasets and runs them through the entire analysis chain. Each
ensemble contains about 7,000 sets of events, where the sets are
constructed to each reproduce {\dzero}'s 2.3~fb$^{-1}$ real
dataset. Signal and background events are sampled from the MC event
sets after all event selection cuts such that the numbers of each
background component match the measured yields, smeared by Poisson
statistics. All systematic uncertainties and their correlations
between background and signal subcomponents are included in the
calculations. The single top quark signal cross section is set at a
different value spanning the range from 2~pb to 10~pb for each
ensemble. For the three discrimination methods and for their
combination, a plot is produced with the measured signal cross section
versus the input signal cross section, and a fit made to the eight
points. For all cases, the slope of the fitted relation is close to
one and the intercept is close to the origin, which shows that the
measured cross section, if it lies in this range, accurately
represents the signal cross section in the data.

\vspace{-0.1in}

\subsection{Single Top Quark Production Cross Sections}

The measured single top quark cross sections are shown in
Table~\ref{cross-sections}. The expected and measured significances of
each measurement are also shown; these are explained in the next
section.

\begin{table}[!h!tbp]
\tbl{Single top quark cross sections and significances from each
analysis. \label{cross-sections}}
{\begin{tabular}{@{}lcccc@{}}
\toprule
         & Single Top    & Uncertainty & \multicolumn{2}{c}{Significance} \\
Analysis & Cross Section &    [\%]     &    Expected     &    Measured    \\
\colrule
{\bf {\dzero}}~~Boosted Decision Trees  & $3.74^{+0.95}_{-0.79}$~pb &                 & $4.3\sigma$  & $4.6\sigma$ \\
~~~~~~~Bayesian Neural Networks         & $4.70^{+1.18}_{-0.93}$~pb &                 & $4.1\sigma$  & $5.4\sigma$ \\
~~~~~~~Matrix Elements                  & $4.30^{+0.99}_{-1.20}$~pb &                 & $4.1\sigma$  & $4.9\sigma$ \\
~~~~~~~{\bf Combination (170~GeV)}   & $\mathbf{3.94\pm0.88}$~{\bf pb} & $\pm22\%$ & $\mathbf{4.5\sigma}$  & ~$\mathbf{5.0\sigma}$
\vspace{0.1in} \\
{\bf CDF}~Boosted Decision Trees       & $2.1^{+0.7}_{-0.6}$~pb    &                 & $5.2\sigma$  & $3.5\sigma$ \\
~~~~~~~~Neural Networks                 & $1.8^{+0.6}_{-0.6}$~pb    &                 & $5.2\sigma$  & $3.5\sigma$ \\
~~~~~~~~Matrix Elements                 & $2.5^{+0.7}_{-0.6}$~pb    &                 & $4.9\sigma$  & $4.3\sigma$ \\
~~~~~~~~Likelihoods                     & $1.6^{+0.8}_{-0.7}$~pb    &                 & $4.0\sigma$  & $2.4\sigma$ \\
~~~~~~~~Likelihoods, s-channel          & $1.5^{+0.9}_{-0.8}$~pb    &                 & $1.1\sigma$  & $2.0\sigma$ \\
~~~~~~~~Combination, lepton+jets        & $2.1^{+0.6}_{-0.5}$~pb    & $+29\%, -24\%$  &              &             \\
~~~~~~~~Neural Networks, {\met}+jets    & $4.9^{+2.6}_{-2.2}$~pb    & $+52\%, -46\%$  & $1.4\sigma$  & $2.1\sigma$ \\
~~~~~~~~{\bf Combination (175~GeV)}  & $\mathbf{2.3^{+0.6}_{-0.5}}$~{\bf pb}    &  $+26\%, -22\%$ & $\mathbf{>5.9\sigma}$ & ~$\mathbf{5.0\sigma}$ \\
~~~~~~~~{\bf Combination (170~GeV)}  & $\mathbf{2.35^{+0.56}_{-0.50}}$~{\bf pb} & $+24\%, -21\%$  &     &     
\vspace{0.1in} \\
{\bf Tevatron Combination (170~GeV)} & $\mathbf{2.76^{+0.58}_{-0.47}}$~{\bf pb}& $+21\%, -17\%$  &     &    
\vspace{0.05in} \\
Theory ($M_{\rm top} = 170$~GeV)        & $3.46\pm0.18$~pb                         & $\pm5\%$        &     &       \\
\botrule
\end{tabular}}
\end{table}

\vspace{-0.1in}

After the two collaborations submitted their independent measurements
for publication, they worked together to combine them into one
Tevatron result.\cite{combination} The systematic uncertainty terms
are classified to map between the two measurements so that
correlations of some terms between the two measurements are properly
taken into account. The combination is calculated using the Bayesian
binned likelihood calculation on all input distributions
simultaneously. The Tevatron combined result is also shown in
Table~\ref{cross-sections}.


\section{Measurement Significance}

The measured significance is defined from the p-value, which is the
probability that the background fluctuated up to give a cross section
measurement at least as large as the measured value. The expected
significance comes from the p-value which is the probability that the
background fluctuated up to give a cross section at least as large as
the standard model theory value. The p-values are converted to
significances in standard deviations ($\sigma$) assuming a Gaussian
distribution. {\dzero} measures these p-values using an ensemble of
about 70 million pseudodatasets, each consisting of only background
events with no signal events, by determining the fraction of
pseudodatasets with a high enough cross section. CDF calculates the
p-values by finding when the quantity $-2 \ln ({\rm Prob}({\rm data}|S
+ B)/{\rm Prob}({\rm data}|B))$ is less in pseudodatasets than in real
data. CDF's pseudodatasets are generated differently to
{\dzero}'s. Instead of sampling MC and multijet background events to
generate each pseudodataset, they sample from the histograms of each
distribution to perform the significance calculation. {\dzero} finds a
measured p-value of $2.5 \times 10^{-7}$ and CDF finds a measured
p-value of $3.1 \times 10^{-7}$. The associated significances are
shown in Table~\ref{cross-sections}. Both experiments have a measured
significance of $5.0\sigma$, meeting the standard to claim
``observation.''


\section{Measuring the CKM Matrix Element $|V_{tb}|$}

The Cabibbo-Kobayashi-Maskawa matrix describes the mixing between
quarks to get from the strong interaction eigenstates to the
weak-interaction ones. The term relating top quarks to bottom quarks
is known as $V_{tb}$. The single top quark production cross section is
proportional to $|V_{tb}|^2$ and can thus be used to measure the
amplitude of $V_{tb}$. To make this measurement, the collaborations
assume the standard model for top quark decay (i.e., mostly to $Wb$
and not much to $Wd$ or $Ws$) and that the $Wtb$ coupling is
left-handed and $CP$-conserving. They do not, however, assume that
there are exactly three quark generations for this measurement. That
is, they do not assume CKM matrix unitarity, unlike measurements of
$|V_{tb}|$ made using top quark decays in {\ttbar}
pairs.\cite{bphysicsheinson} The measurements include uncertainties
from the $tb$+$tqb$ theory cross section as well as those included in
the cross section measurement. The theory cross section uncertainty
from the top quark mass uncertainty is 4.2\%, with 3.0\% from the
PDFs, 2.4\% from the factorization scale, and 0.5\% from the strong
coupling constant $\alpha_s$. Two measurements of $|V_{tb}|$ are made:
the first does not constrain the strength of the left-handed scalar
coupling constant $f_1^L$ (where a nonnegative prior is used, with no
upper bound), and the second sets $f_1^L = 1$ (when the prior is
bounded between zero and one). The results of this measurement from
{\dzero}, CDF, and the Tevatron combination, are shown in
Table~\ref{Vtb}.

\begin{table}[!h!tbp]
\tbl{Measurements of the CKM matrix element $|V_{tb}|$.
\label{Vtb}}
{\begin{tabular}{@{}lcc@{}}
\toprule
Experiment & $|V_{tb}|f_1^L$ & $|V_{tb}|$ ($f_1^L = 1$) \\
and Theory &                 &  95\% CL                 \\
\colrule
{\dzero} ~~~$M_{\rm top} = 170$~GeV  & ~~~$1.07 \pm 0.12$~~~ & $0.78 < |V_{tb}| \le 1$ \\
~~~~~~~~~Kidonakis 2006              &                       &
\vspace{0.1in} \\
CDF ~~$M_{\rm top} = 175$~GeV        &                       & $0.71 < |V_{tb}| \le 1$ \\
~~~~~~~~~Harris {\it et al.} 2002   &                       &
\vspace{0.1in} \\
Tevatron Combination                 &                       &                         \\
~~~~~~~~~$M_{\rm top} = 170$~GeV     &                       & $0.77 < |V_{tb}| \le 1$ \\
~~~~~~~~~Kidonakis 2006              &                       &                         \\
\botrule
\end{tabular}}
\end{table}


\section{Separate s-Channel and t-Channel Measurements}

\begin{figure}[!h!tbp]
\begin{center}
\epsfig{file=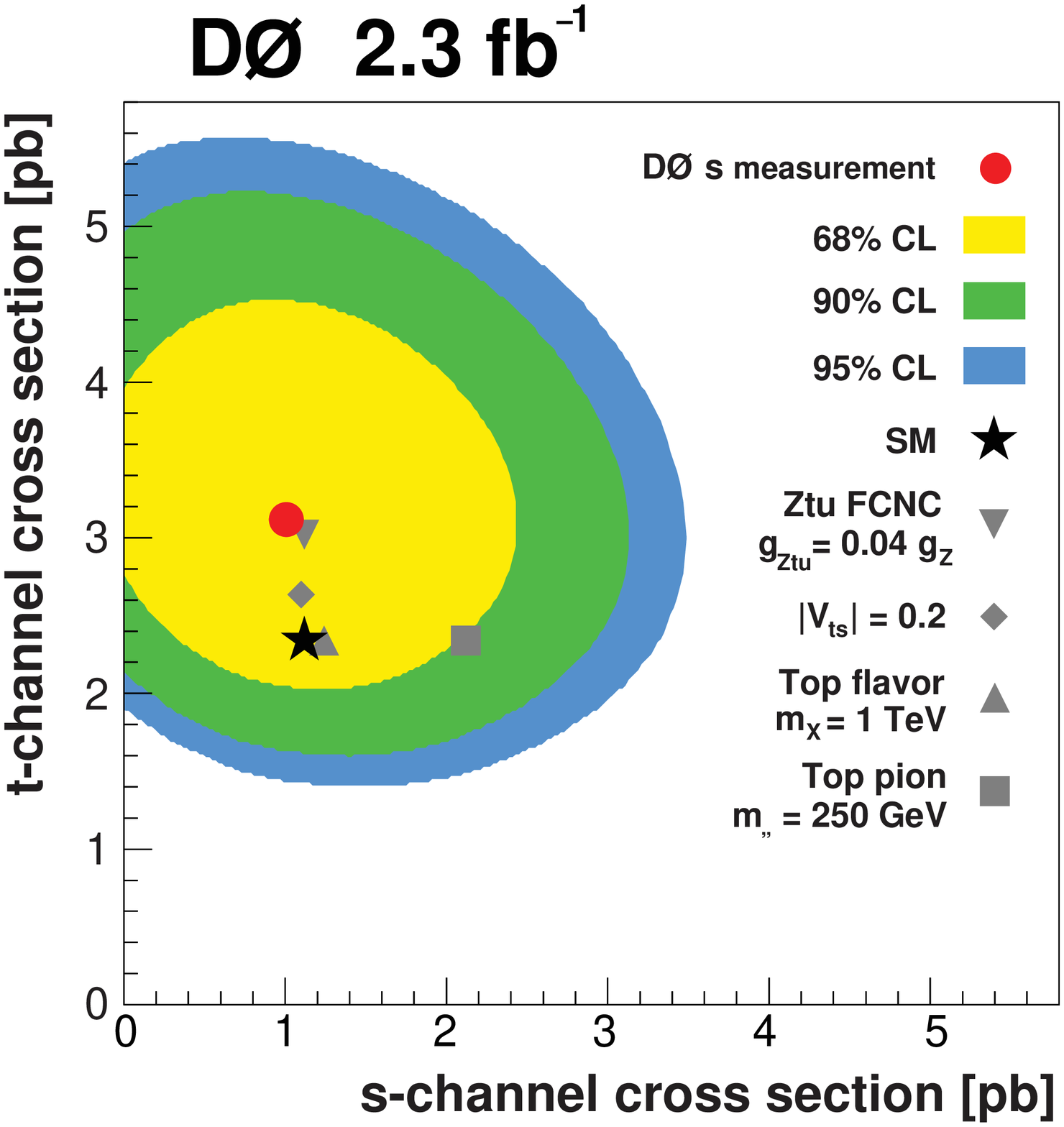,width=2.3in}
\hspace{0.2in}
\epsfig{file=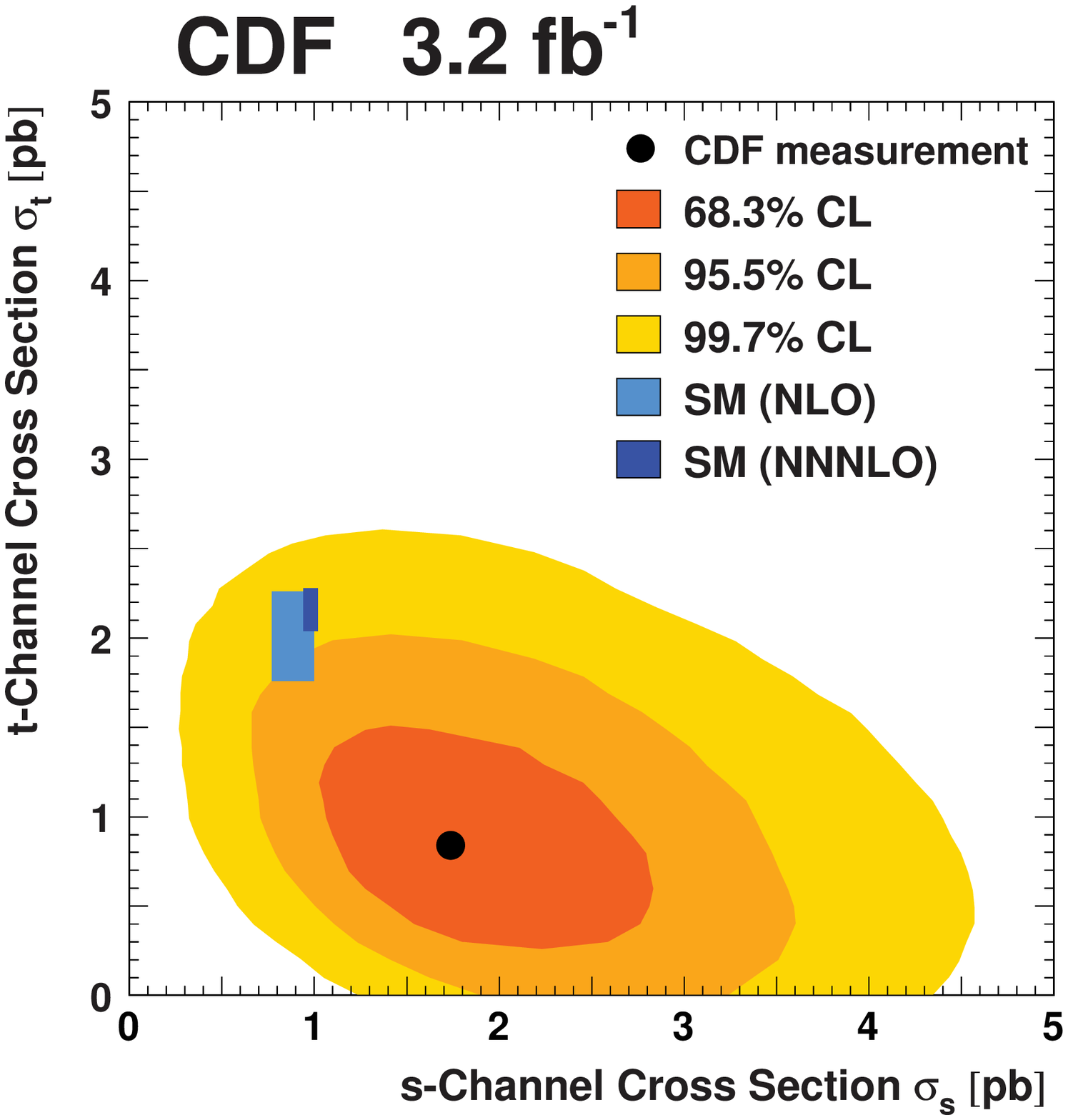,width=2.3in}
\end{center}
\vspace{-0.1in}
\caption[2dplot]{Plots showing the separate s-channel $tb$ and
t-channel $tqb$ cross section measurements from (a)~{\dzero} and
(b)~CDF, together with theory values and some beyond-the-SM model
predictions. {\dzero}'s measurements are for $M_{\rm top} = 170$~GeV
and CDF's for 175~GeV. The theory cross sections shown are Kidonakis
2006 for {\dzero}, Harris {\it et al.} 2002 for CDF (``NLO'') and
Kidonakis 2006 for CDF (``NNNLO''). \label{schan-tchan}}
\end{figure}

Both collaborations have made measurements of the s-channel $tb$ and
t-channel $tqb$ single top quark cross sections separately. {\dzero}
retrains the three sets of discriminants with just t-channel single
top as the signal, instead of $tb$+$tqb$ as in the observation
analysis.\cite{singletopdzerotchan} CDF uses measurements obtained
during their main analysis. Neither of these measurements assume the
SM ratio for the s-channel and t-channel cross sections, unlike in the
observation analysis. The results are shown in
Fig.~\ref{schan-tchan}. {\dzero}'s t-channel cross section measurement
has a significance of $4.8\sigma$.

\vspace{-0.3in}

\section{Summary}

The {\dzero} and CDF collaborations have searched large Tevatron
datasets and observed single top quark production for the first time,
with $5\sigma$ significance for each of the measurements. The measured
cross sections are consistent with NLO theory predictions. The
analyses have been improved in many ways to achieve this, in
particular over the years of the search:

\vspace{0.075in}
\noindent {\bf {\dzero}'s Innovations}
\vspace{-0.05in}

\begin{itemlist}

\item Next-to-leading order simulation of signals with full spin
information included

\item Very loose kinematic cuts and use of all possible triggers to
select more signal-like data and increase signal acceptance

\item Multivariate techniques: BDTs, BNNs, and MEs used to improve
signal-background separation

\item Very large number of variables, including many not used at the
Tevatron before such as jet widths, to separate signal from background

\item Rebinning the discriminant outputs to ensure no bin has data or
expected signal and no background events, which stabilizes the cross
section measurement

\end{itemlist}

\noindent {\bf CDF's Innovations}
\vspace{-0.05in}

\begin{itemlist}

\item Including data with no identified lepton to extend the signal
acceptance

\item Jet flavor separator variable to discriminate $b$~jets from
mistagged charm, light quark, and gluon ones after $b$~tagging

\item Combining different measurements using the NEAT algorithm
optimized to maximize the expected signal significance

\item Binned likelihood fit to a discriminating variable shape to
improve the measurement sensitivity

\end{itemlist}

As the reader can see from these lists and the previous analysis
descriptions, each collaboration has learned from the innovations of
the other one, and the outcome is a deep understanding of the signals
and multicomponent backgrounds in many analysis channels, with
powerful new analysis techniques developed to extract a small signal
from a very large background. Many of these techniques are now being
applied to the search for the Higgs boson at the Tevatron, and the
single top datasets are providing a unique place in which to test
various aspects of the standard model and search for new
physics.\cite{newphysics1}$^-$\cite{newphysics9}


\section*{Acknowledgments}

I have worked with a large number of people from the {\dzero}
collaboration on this measurement and previous iterations over the
past 15 years. Cecilia Gerber and Reinhard Schwienhorst co-led the
team with me that made the observation measurement, and Ar\'{a}n
Garc\'{\i}a-Bellido co-led with me the evidence analysis in
2006. Fourteen students have obtained their Ph.D.s on the research at
{\dzero}, seven students have graduated with master's degrees, nine
Ph.D.  students are still in the pipeline, and six postdocs have
become tenure-track professors or lab staff scientists. It has been a
pleasure to collaborate with all of the single top working group
members and I have learned a lot from them. I have also enjoyed
interactions with members of CDF's single top working group over the
years, and with many theorists who generously shared their
perspectives and insights on single top quark physics. This paper is
based on a plenary talk I presented at the 21st Rencontres de Blois:
Windows on the Universe, Blois, France, June 2009, and associated
proceedings paper.\cite{bloisheinson} For this review paper, Reinhard
Schwienhorst and Liang Li from {\dzero} and Tom Junk and Jan L\"{u}ck
from CDF helped answer my questions. This research has been supported
by grants from the U.S. Department of Energy's Office of Science.


\end{document}